\documentclass[aps,prb,reprint,twocolumn,superscriptaddress,amsmath,amssymb,floatfix]{revtex4-2}
\usepackage[utf8]{inputenc}
\usepackage[T1]{fontenc}
\usepackage{lmodern}
\usepackage{graphicx}
\usepackage{bm}
\usepackage{booktabs}
\usepackage{longtable}
\usepackage{makecell}
\usepackage{multirow}
\usepackage{tabularx}
\usepackage{adjustbox}
\usepackage[table]{xcolor}
\usepackage[hidelinks]{hyperref}
\newcommand{\ms}{\mathrm{mSCAN}}
\newcommand{\hx}{\mathrm{hole}}
\newcommand{\Sr}{\mathrm{SR}}
\newcommand{\Lr}{\mathrm{LR}}
\newcommand{\Fx}{F_x}

\begin{document}

\newcommand{\szs}[1]{\textcolor{blue}{\textbf{[#1]}}}
               
\title{
Exchange-hole model for modified SCAN meta-GGA functional: Real-space constraints and screened hybrid exchange
}

\author{Subrata Jana}
\email{subrata.niser@gmail.com, subrata.jana@umk.pl}
\affiliation{Institute of Physics, Faculty of Physics, Astronomy and Informatics, Nicolaus Copernicus University in Toru\'n, ul. Grudzi\k{a}dzka 5, 87-100 Toru\'n, Poland}
 \affiliation{Institute of Advanced Studies, Nicolaus Copernicus University in Toru\'{n}, ul. Wile\'{n}ska 4, 87-100 Toru\'{n}, Poland} 

\author{Szymon \'Smiga}
\affiliation{Institute of Physics, Faculty of Physics, Astronomy and Informatics, Nicolaus Copernicus University in Toru\'n, ul. Grudzi\k{a}dzka 5, 87-100 Toru\'n, Poland}
 \affiliation{Institute of Advanced Studies, Nicolaus Copernicus University in Toru\'{n}, ul. Wile\'{n}ska 4, 87-100 Toru\'{n}, Poland} 

\date{\today}
\begin{abstract}
Screened hybrid functionals require a separation of semilocal exchange into short- and long-range contributions, which cannot be uniquely determined from the exchange-energy density alone. Here, we develop an analytic, reverse-engineered exchange-hole model for a modified strongly constrained and appropriately normed (mSCAN) semilocal exchange functional. The model explicitly depends on the reduced density gradient and the orbital kinetic-energy indicator and satisfies the on-top value, particle-normalization condition, and exchange-energy moment. Its analytic screened moment provides the short-range semilocal exchange contribution for constructing HSE-mSCAN, a screened meta-GGA hybrid functional employing 10\% short-range Fock exchange. We examine the real-space and wave-vector-resolved properties of the exchange hole and assess its implications for the exchange--correlation response of the uniform electron gas. The resulting framework establishes a route to exchange-hole-based range separation of meta-GGA exchange and provides a foundation for developing dielectric-dependent, long-range-corrected, and double-hybrid functionals.
\end{abstract}

\maketitle

\section{Introduction}

Kohn--Sham (KS) density-functional theory (DFT) describes the ground-state density of an interacting electron system through an auxiliary system of noninteracting electrons~\cite{hohenberg1964,kohn1965self,levy1979,lieb1983}. Its predictive accuracy depends largely on the approximation to the exchange--correlation (XC) energy and on the property of interest~\cite{burke2012perspective,kaplan2023constraints}. An approximation that yields accurate equilibrium structures and binding energies can still give substantial errors in the fundamental gap, $E_{\mathrm{g}}=I-A$, where $I$ is the ionization energy and $A$ the electron affinity. In exact KS theory with a multiplicative potential, the fundamental gap equals the KS eigenvalue gap plus the derivative discontinuity of the XC potential~\cite{pplb1982,perdewlevy1983,sham1983}. The generalized KS (GKS) framework also admits nonmultiplicative XC operators and underlies the meta-GGA and hybrid calculations considered here~\cite{seidl1996,perdew2017gaps}.

The strongly constrained and appropriately normed (SCAN) meta-generalized gradient approximation (meta-GGA) uses the density $n$, its gradient, and the orbital kinetic-energy density $\tau$ to describe a range of bonding environments within a semilocal functional~\cite{scan2015,sun2016}. Its dimensionless iso-orbital indicator $\alpha$ distinguishes regions dominated by a single spatial orbital from those characteristic of slowly varying densities or overlapping closed shells. Subsequent regularizations have improved numerical stability while preserving much of SCAN's constraint-based construction~\cite{bartok2019,furness2020,r2scancorrection}. Challenges remain for self-interaction and fractional-charge errors and for the prediction of band gaps~\cite{cohen2008,mori2008frac,jana2018sie}. SCAN also overestimates the magnetic moments of the itinerant ferromagnets Fe, Co, and Ni~\cite{ekholm2018scan}.

Hybrid functionals replace a fraction of semilocal exchange with nonlocal Fock exchange~\cite{becke1993,pbe0}. Range separation allows this fraction to differ between the short-range (SR) and long-range (LR) parts of the electron--electron interaction~\cite{iikura2001,toulouse2004,yanai2004,vydrov2006}. The Heyd--Scuseria--Ernzerhof (HSE) family, widely used for periodic solids, confines Fock exchange to SR~\cite{hse2003,hse2006erratum,krukau2006,jana2019screened,jana2020screened,jana2020improved,jana2019long}. Dielectric-dependent (DD) hybrids relate the exchange fraction to the dielectric response of the material, either globally or through a range-dependent prescription~\cite{marques2011,skone2014,skone2016,chen2018}. Recent examples combine dielectric screening with meta-GGA ingredients~\cite{jana2025hybrid,jana2025twoD,riemelmoser2026}. For a screened hybrid with a specified interaction kernel, a separate question remains: how should the semilocal exchange of the parent functional be partitioned between short and long range?

The exchange hole provides a physical basis for this partition. Around a reference electron, the hole describes a deficit of same-spin density that integrates to one missing electron. Its Coulomb-weighted moment determines the exchange energy. Different radial hole profiles can yield the same Coulomb moment yet give different contributions when weighted by an SR kernel. The parent exchange-energy density alone therefore leaves the range separation undetermined. Models of the exchange hole draw on its SR expansion, the uniform-electron-gas limit, and other exact constraints~\cite{beckeroussel1989,constantin2009hole,perdew1996hole,ernzerhof1998hole}. Reverse engineering uses the exchange-energy density of a chosen functional to constrain the model, providing an analytical route to SR exchange for GGAs~\cite{hjs2008}. At the meta-GGA level, the hole can also retain information about the local orbital environment through its dependence on $\tau$~\cite{constantin2006hole,tpsshole,jana2018pccp,jana2019lr}.

In this work, we construct a reverse-engineered exchange hole that depends explicitly on both the reduced density gradient $s$ and the iso-orbital indicator $\alpha$. To accommodate the chosen radial form and SR curvature, we omit the large-gradient damping factor in SCAN exchange. This modification defines the parent exchange functional, denoted mSCAN, and changes its large-gradient and dimensional-crossover behavior. We impose the particle sum rule, the on-top value, and the exchange-energy moment of mSCAN algebraically. We assess pointwise nonpositivity of the resulting hole numerically.

Integrating the mSCAN hole against complementary SR and LR kernels gives the two semilocal exchange contributions used to construct HSE-mSCAN. This screened hybrid replaces 10\% of SR mSCAN exchange with Fock exchange, retains full LR mSCAN exchange, and uses modified SCAN correlation. We evaluate this fixed-parameter functional against established semilocal and hybrid approximations. In particular, the comparison with HSE06 reflects differences in the semilocal exchange and correlation functionals as well as in the fraction of SR Fock exchange. For solids, we examine lattice constants, bulk moduli, and cohesive energies of 46 cubic solids and band gaps of 76 solids, with additional tests for oxides and polymorphs. We also compare the relative energies of water-hexamer isomers.

\section{Exchange-hole construction}\label{sec:model}
\subsection{Exchange target and radial moments}
For a spin-unpolarized density one can define $k_F=(3\pi^2 n)^{1/3}$, $s=|\nabla n|/(2k_Fn)$, $p=s^2$, and $\alpha=(\tau-\tau_W)/\tau_{\rm unif}$, where $\tau=\frac12\sum_i f_i|\nabla\psi_i|^2$, $\tau_W=|\nabla n|^2/(8n)$, and $\tau_{\rm unif}=3k_F^2n/10$. Here $f_i$ are orbital occupations, including spin degeneracy for spatial orbitals. For simplification, we propose the modified exchange enhancement as, 
\begin{align}
\Fx^{\ms}(p,\alpha)&=h_1(p,\alpha)
 +f_x(\alpha)[1.174-h_1(p,\alpha)],\label{eq:fm}\\
h_1(p,\alpha)&=1+\frac{k_1x(p,\alpha)}{k_1+x(p,\alpha)},\nonumber\\
x(p,\alpha)&=\mu p+b_4p^2 e^{-|b_4|p/\mu}\nonumber\\
&\quad+\bigl[b_1p+b_2(1-\alpha)e^{-(1-\alpha)^2/2}\bigr]^2.\nonumber
\end{align}
Here $k_1=0.065$, $\mu=10/81$, $b_2=(5913/405000)^{1/2}$, $b_1=511/(27000b_2)$, and $b_4=\mu^2/k_1-1606/18225-b_1^2$, with
\begin{equation}
 f_x(\alpha)=\begin{cases}
 e^{-0.667\alpha/(1-\alpha)},&0\le\alpha<1,\\
 0,&\alpha=1,\\
 -1.24e^{0.8/(1-\alpha)},&\alpha>1.
 \end{cases}\label{eq:fxinterp}
\end{equation}
The original SCAN exchange is $\Fx^{\rm SCAN}=g_x(s)\Fx^{\ms}$ with $g_x(s)=1-\exp[-4.9479/\sqrt{s}]$ for $s>0$, and $g_x(0)=1$~\cite{scan2015}. Thus, mSCAN modifies the large-gradient behavior and the hydrogen norm~\cite{supplement}.

Now, using the spherically averaged hole $\bar h_x(\mathbf r,u)=n(\mathbf r)J(p,\alpha,y)$, with $y=k_F(\mathbf r)u$ and $u=|\mathbf r-\mathbf r'|$, the exchange-energy functional is defined as,
\begin{equation}
 E_x=2\pi\int d^3r\,n(\mathbf r)\int_0^\infty du\,u\,\bar h_x(\mathbf r,u).
\end{equation}
Here, the shape function $J$ is prescribed through the similar model used in Henderson–Janesko–Scuseria (HJS)~\cite{hjs2008} and MGGAC~\cite{jana2022}, which satisfies the well-known hole constraints~\cite{hjs2008,tpsshole,jana2022}
\begin{subequations}\label{eq:moments}
\begin{align}
J(p,\alpha,0)&=-\tfrac12,\label{eq:ontop}\\
\int_0^\infty dy\,y^2J(p,\alpha,y)&=-\tfrac{3\pi}{4},\label{eq:norm}\\
\int_0^\infty dy\,yJ(p,\alpha,y)&=-\tfrac98\Fx^{\ms}(p,\alpha),\label{eq:energy}\\
J(p,\alpha,y)&\le0.\label{eq:negative}
\end{align}
\end{subequations}
The first three constraints can be enforced algebraically. Their satisfaction does not establish the pointwise inequality in Eq.~\eqref{eq:negative}.

\subsection{Hole ansatz and determination of parameters}
We consider the following non-oscillatory model for the exchange hole shape function
\begin{align}
J(p,\alpha,y)={}&-\frac{9e^{-Hy^2}}{4y^4}(1-e^{-Ay^2})\nonumber\\
&+\left(\frac{9A}{4y^2}+B+Uy^2+Vy^4\right)e^{-(D+H)y^2},\label{eq:hole}
\end{align}
where $H$, $U$, and $V$ depend on $(p,\alpha)$ with $A=0.757211$, $D=0.6096500002334342$, and $B=-0.1063637666009172$. The fixed constants reproduce the specified non-oscillatory uniform-gas hole, and the inverse powers of $y$ cancel in the small-separation expansion.

The shape functional $J(p,\alpha,y)$ can also be defined through the curvature of the shape function $L$ as, 
\begin{equation}
 J(p,\alpha,y)=-\frac12+\frac{L(p,\alpha)}{2}y^2+O(y^4),
 \label{eq:small-y}
\end{equation}
That satisfies $\frac{\partial J}{\partial y}|_{y=0}=L(p,\alpha)$, another well known constraints on the shape function~\cite{jana2022}.

Thus Comparison with Eq.~\eqref{eq:hole} gives
\begin{equation}
U=C_0+\tfrac12(L-\tfrac15-H),\label{eq:U}
\end{equation}
with $C_0=-0.1186486691319622$. 
In a similar way as defined in MGGAC~\cite{jana2022} and HJS hole~\cite{hjs2008}, we set $\lambda=D+H$ and $\eta=A+H$ and
% \szs{for me these two relations $\lambda=D+H$, $\eta=A+H$ are not clear and I do not understand why we assume that?}
% {\textcolor{red}{This is conventional way of defining the terms also used in HJS and MGGAC paper}}
applying the particle sum rule yields
\begin{align}
V={}&\frac{6A\lambda^3}{5}
\left(\frac{2\sqrt{\lambda}}
{\sqrt{H}+\sqrt{\eta}}-1\right)
-\frac{4\sqrt{\pi}\lambda^{7/2}}{5}
\nonumber\\
&-\frac{4B\lambda^2}{15}
-\frac{2\lambda U}{5}.
\label{eq:V}
\end{align}
Consequently, only $H$ remains to be determined from the
exchange-energy condition. The enhancement factor generated
by the exchange hole is
\begin{align}
\mathcal{F}^{\hx}(H;L)
={}&\frac{A}{15}
\left(31-\frac{32\sqrt{\lambda}}
{\sqrt{H}+\sqrt{\eta}}\right)
-\frac{28B}{135\lambda}
\nonumber\\
&+\frac{32\sqrt{\pi\lambda}}{45}
-\frac{4U}{45\lambda^2}
+H\ln\frac{H}{\lambda}
-\eta\ln\frac{\eta}{\lambda},
\label{eq:Fhole}
\end{align}
where $H\ln(H/\lambda)=0$ at $H=0$.
We determine $H\geq 0$ by solving
$\mathcal{F}^{\hx}(H;L)=\Fx^{\ms}(p,\alpha)$,
and subsequently obtain $U$ and $V$ from
Eqs.~\eqref{eq:U} and \eqref{eq:V}, respectively.

%\begin{align}
%N_0(H)=\sqrt{\pi}\biggl[&-\frac{9A}{4(\sqrt H+\sqrt\eta)}
%+\frac{9A}{8\sqrt\lambda}\nonumber\\
%&+\frac{B}{4\lambda^{3/2}}\biggr],\qquad
% R_N(H)=-\frac{3\pi}{4}-N_0(H)~,\label{eq:N0}
%\end{align}
%and applying the particle sum rule $V$  fixes to
%\begin{equation}
% V=\frac{16\lambda^{7/2}}{15\sqrt\pi}R_N(H)-\frac25\lambda U.\label{eq:V}
%\end{equation}
%Consequently, only $H$ remains to be determined from the exchange-energy condition. Now introducing
%\begin{align}
%M_1(H)=-\frac98\biggl[A-\frac{4B}{9\lambda}
%+H\ln\frac H\lambda-\eta\ln\frac\eta\lambda\biggr],\label{eq:M1}
%\end{align}
%with $H\ln(H/\lambda)=0$ at $H=0$, the generated enhancement function becomes,
%\begin{equation}
%\mathcal F^{\hx}(H;L)=-\frac89\left[M_1(H)+\frac{U}{10\lambda^2}
%+\frac{16\sqrt\lambda}{15\sqrt\pi}R_N(H)\right].\label{eq:Fhole}
%\end{equation}
%We solve $\mathcal F^{\hx}(H;L)=\Fx^{\ms}(p,\alpha)$ for $H\ge0$, then obtain $U$ and $V$ from Eqs.~\eqref{eq:U} and \eqref{eq:V}.

Now the only remaining quantity is $L(p,\alpha)$. To construct this, we use the following starting curvature:
\begin{align}
 L_b&=\frac15-\frac{4p}{9(1+p)}
+w(\alpha)\frac{49p+9}{90(1+p)(2+p)},\label{eq:Lb}\\
 L_E&=\frac15+\frac{1-\Fx^{\ms}(p,\alpha)}{K},
 \quad K=\frac2{45D^2},\label{eq:LE}\\
 L(p,\alpha)&=\frac12\bigl[L_b+L_E+\sqrt{(L_b-L_E)^2+\epsilon^2\Xi^4}\bigr],\nonumber\\
\Xi&=\frac{p+(\alpha-1)^2}{1+p+(\alpha-1)^2},
\quad\epsilon=0.01,\label{eq:L}
\end{align}
where $w(\alpha)=\exp[-0.667\alpha/(1-\alpha)]$ for $0\le\alpha<1$ and $w=0$ otherwise. The dimensionless variable $\Xi$ controls the smooth interpolation
between the two model curvatures. The rounded maximum in Eq.~\eqref{eq:L} ensures $L\ge L_E$, which provides a nonnegative-root existence condition for the energy equation. %It does not ensure that the root is unique or that the resulting hole is nonpositive. 
One may note that at $(p,\alpha)=(0,1)$, $H=0$ and $L=1/5$ recover the chosen uniform-gas moments and curvature. %The full exact orbital-hole curvature, which also contains density-Laplacian information, is not enforced by this ansatz. 

\subsection{Why the original SCAN limit is not reproduced}\label{sec:scandiff}
% For fixed $\alpha$ value, the SCAN enhancement factor factor obeys $g_x(s)\sim4.9479s^{-1/2}$ as $s\to\infty$. Reproducing this diminishing energy moment with a normalized negative hole requires its missing charge to extend to increasingly large electron separations. Forcing the original SCAN target into the present radial ansatz requires a large change in the model curvature and can generate positive values of $J$, which is a clear violation of constraint Eq.~\eqref{eq:negative}. We therefore use mSCAN, omitting $g_x(s)$, while retaining the same explicit $s$- and $\alpha$-dependent hole construction. This is a modification of the functional's large-gradient behavior and dimensional-crossover properties, rather than a demonstration that a normalized nonpositive hole for original SCAN is impossible. 

% As a diagnostic of the model limitation, we repeated exactly the same procedure after replacing the mSCAN energy target by $\Fx^{\rm SCAN}=g_x(s)\Fx^{\ms}$, without changing the ansatz or curvature rule. The forced-SCAN calculation gave positive holes% at 1171 of the 5084 parameter pairs
% (Fig.~\ref{fig:sign}). At $s=100$ and $\alpha=1$, the forced SCAN fit has $\Fx=0.415671$ and $J_{\max}\simeq+0.597$ near $y=1.07$; the mSCAN target has $\Fx=1.065$ and a nonpositive sampled profile. The forced fit is not a proposed model of original SCAN. It demonstrates why matching the energy moment and normalization alone is an insufficient sign test.

At fixed $\alpha$, the large-$s$ decay of the SCAN exchange enhancement
factor is governed by $g_x(s)\sim 4.9479\,s^{-1/2}$. For a normalized,
nonpositive exchange hole to reproduce the resulting vanishing energy
moment, its charge deficit must shift to increasingly large scaled
electron separations. Within the present radial ansatz, matching the
SCAN energy target can require a substantial change in the hole
curvature and produce regions where $J>0$, in violation of
Eq.~\eqref{eq:negative}. We therefore use the mSCAN energy target,
which omits $g_x(s)$, while retaining the explicit dependence of the
hole on both $s$ and $\alpha$. This choice modifies the large-gradient
behavior of the functional and its dimensional-crossover properties.
The limitation identified here is specific to the present hole
construction.

\begin{figure*}[t]
\centering
\includegraphics[width=0.88\textwidth]{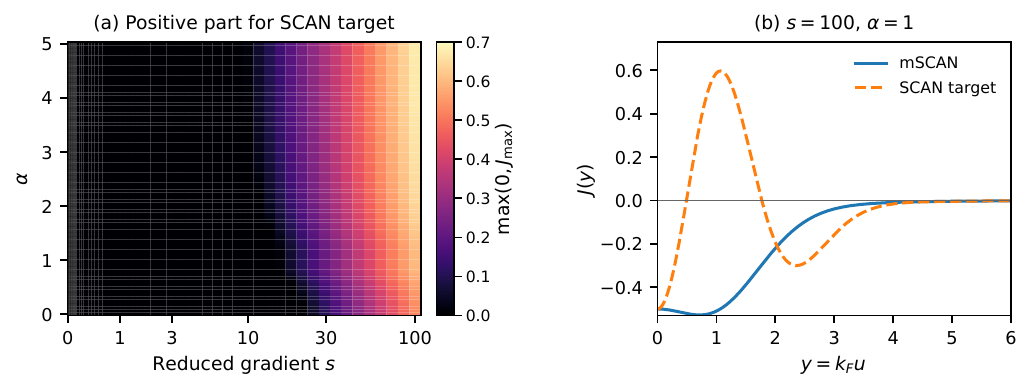}
\caption{Numerical sign diagnostic for the same radial ansatz. (a) Positive part of the maximum sampled hole, $\max[0,\max_{0\le y\le150}J]$, when the energy target is replaced by the original SCAN (logarithmically compressed $s$ axis). The mSCAN target has no positive values above $10^{-6}$ on the sampled grid. (b) Resulting holes at $s=100$, $\alpha=1$. The SCAN-target curve violates $J\le0$ despite satisfying the imposed moments. }
%Neither panel is a global existence or uniqueness proof.}
\label{fig:sign}
\end{figure*}

To isolate the effect of the energy target, we repeated the fit with
$\Fx^{\rm SCAN}=g_x(s)\Fx^{\ms}$ using the same radial ansatz and
curvature rule. For some $(s,\alpha)$ pairs, the fitted hole becomes
positive over part of the radial range (Fig.~\ref{fig:sign}). For
example, at $s=100$ and $\alpha=1$, the SCAN target gives
$\Fx=0.415671$, and the fitted hole reaches $J_{\max}\simeq +0.597$
near $y=1.07$. At the same $(s,\alpha)$, the mSCAN target gives
$\Fx=1.065$, and the hole is nonpositive at every sampled separation.
This diagnostic shows that a hole can satisfy both normalization and
the energy moment while violating pointwise nonpositivity.

\section{Real-space constraints and screened exchange}\label{sec:realspace}
\subsection{Real-space hole profiles}\label{sec:gridtest}

\begin{figure*}[t]
\centering
\includegraphics[width=1\textwidth]{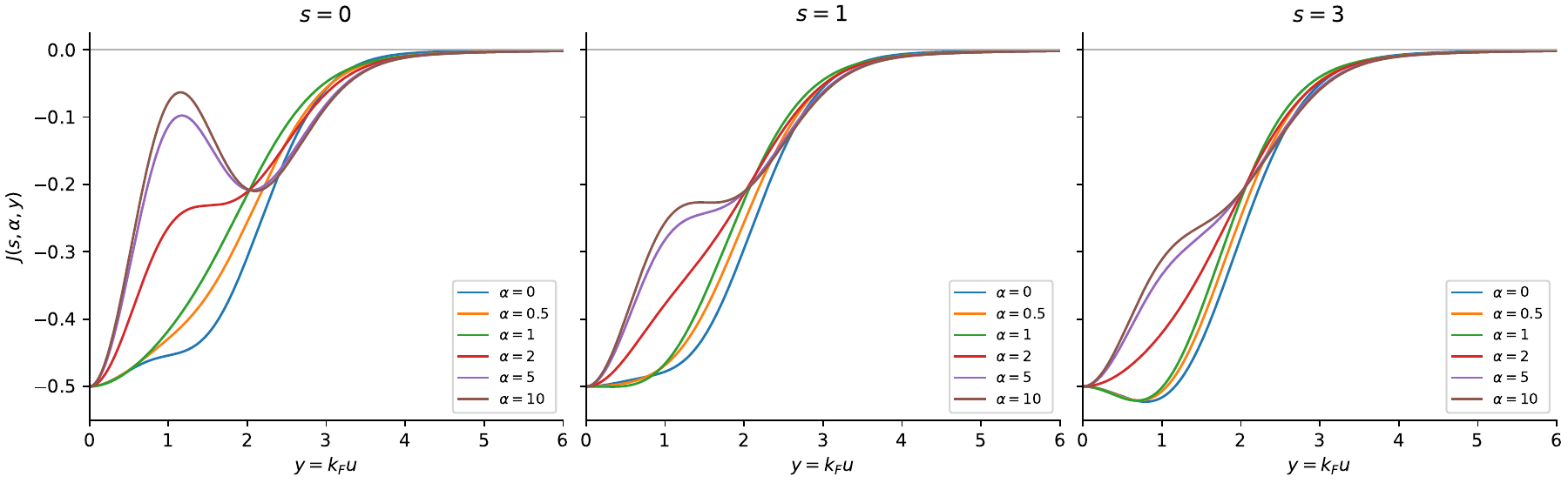}
\caption{Radial exchange-hole profiles $J(p=s^2,\alpha,y)$ for $s=0$, 1, and 3 and $\alpha=0$, 0.5, 1, 2, 5, and 10. The separation is $y=k_Fu$, with the local $k_F$. Each curve uses the mSCAN target and the same root prescription. All displayed values are nonpositive.}
\label{fig:profile}
\end{figure*}

Figure~\ref{fig:profile} shows the radial exchange-hole profiles
$J(p,\alpha,y)$ for $p=s^2$, with $s=0$, $1$, and $3$, and for several
values of the orbital kinetic-energy indicator $\alpha$. The scaled
separation $y=k_Fu$ expresses the electron-electron distance $u$ in
units of the inverse local Fermi wave vector, $k_F^{-1}$. All profiles
satisfy the imposed on-top condition $J(p,\alpha,0)=-1/2$, remain
nonpositive throughout the plotted range, and approach zero at large $y$.

The profiles are most sensitive to $\alpha$ at intermediate separations.
For small $\alpha$, the hole remains appreciably negative over a broader
range of $y$. As $\alpha$ increases, a shoulder develops around
$y=1$--$2$, reflecting a redistribution of the exchange hole at
intermediate separations. The shoulder is most pronounced at $s=0$ and
weakens as $s$ increases. At $s=3$, the profiles approach zero more
smoothly, although their dependence on $\alpha$ remains visible.
The present construction thus extends the earlier $\alpha$-dependent
exchange-hole model~\cite{jana2022} by allowing density inhomogeneity,
through $s$, to modify the radial profile even at fixed orbital
character, as measured by $\alpha$.

\subsection{Wave-vector analysis of the exchange surface energy}
\label{sec:wavevector}

\begin{figure*}[t]
\centering
\includegraphics[width=0.70\textwidth]{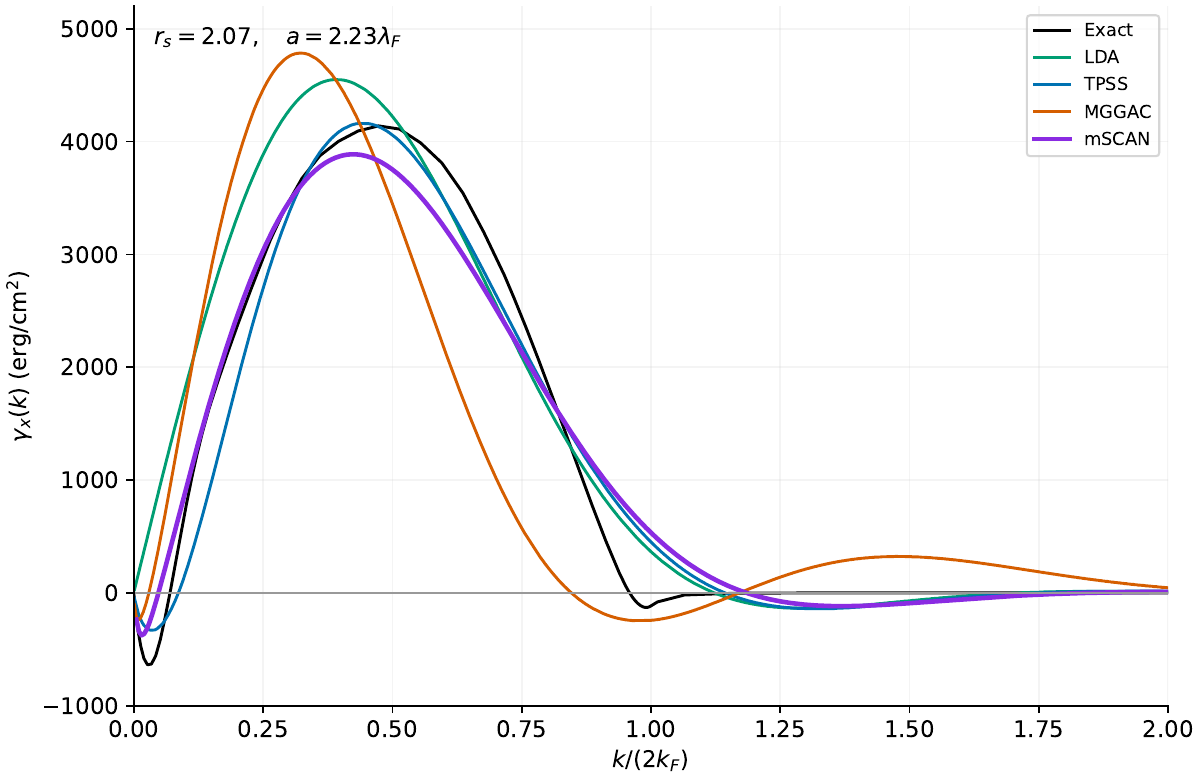}
\caption{Wave-vector-resolved exchange surface-energy spectrum
$\gamma_x(k)$ for a neutral jellium slab with
$r_s=2.07$ and positive-background thickness
$a=2.23\lambda_F$. The horizontal axis is the
normalized wave vector $k/(2k_F)$, where
$k_F=(3\pi^2\bar n)^{1/3}$ is the bulk Fermi
wave vector. The calculated mSCAN corresponds to the exchange-hole
model evaluated using fixed LDA/PW92 slab orbitals.
The exact-exchange, LDA, TPSS, and MGGAC spectra
are extracted from Ref.~\cite{jana2022} and are
shown for comparison.}
\label{fig:wv}
\end{figure*}

The exchange-hole model can be further assessed by examining its wave-vector-resolved contribution to the jellium surface exchange energy. This quantity provides a more detailed test of the spatial distribution of the exchange hole than the integrated exchange energy, as it reveals how different wave-vector components contribute to the exchange energy associated with surface formation. This wave-vector analysis has been established in earlier studies of jellium surfaces~\cite{langreth1977,pitarke2006surface,tpsshole,jana2022}. To perform this analysis, we consider a neutral jellium slab, in which the positive ionic background is represented by a uniform charge distribution. The slab is infinite in the two directions parallel to the surface, while the electron density varies along the surface-normal direction, $z$.

Following Ref.~\cite{jana2022}, we consider a slab with $r_s=2.07$ and a positive-background thickness $a=2.23\lambda_F$. Here, $r_s$ is the Wigner--Seitz density parameter, defined by $4\pi(r_s a_0)^3\bar n/3=1$, where $a_0$ is the Bohr radius and $\bar n$ is the uniform positive-background density. The corresponding bulk Fermi wave vector and Fermi wavelength are given by
\begin{equation}
k_F=(3\pi^2\bar n)^{1/3},
\qquad
\lambda_F=\frac{2\pi}{k_F}.
\end{equation}
% \szs{Write here why you we need $n(z)$. This look bit ad hook definition. } {\textcolor{red}{This is how inside the code it was used. Actually electron density varies along the durface-normal direction, z}}
The electron density $n(z)$ and the occupied
Kohn--Sham orbitals are obtained from a
one-dimensional LDA/PW92 slab calculation
~\cite{pw1992,jana2022}. The proposed exchange
hole is subsequently evaluated using these fixed
density and orbitals. Thus, the present analysis
is a non-self-consistent test of the exchange-hole
model rather than a screened-hybrid slab calculation.

We denote the electron--electron separation by
$R=|\mathbf r-\mathbf r'|$, where $\mathbf r$ is
the position of the reference electron and
$\mathbf r'$ is the position of another electron.
The dimensionless separation is defined as
$u=k_FR$, where $k_F$ is the bulk Fermi wave
vector. Introducing the normalized density
$\rho(z)=n(z)/\bar n$, the local Fermi wave vector
becomes
\begin{equation}
k_F^{\mathrm{loc}}(z)
=k_F\rho(z)^{1/3}.
\end{equation}
Consequently, the argument of the exchange-hole
shape function is
\begin{equation}
y=k_F^{\mathrm{loc}}(z)R
=u\rho(z)^{1/3}.
\end{equation}
The local quantities $p(z)$ and $\alpha(z)$
are the squared reduced density gradient and
the kinetic-energy-density indicator, respectively,
defined as
\begin{equation}
p(z)=\left[
\frac{|dn(z)/dz|}
{2k_F^{\mathrm{loc}}(z)n(z)}
\right]^2,
\end{equation}
and
\begin{equation}
\alpha(z)=
\frac{\tau(z)-\tau^W(z)}
{\tau^{\mathrm{unif}}(z)}.
\end{equation}
% \szs{Do we need to define all of these $z$ dependent quantities? } {\textcolor{red}{This is how inside the code it was used. Actually electron density varies along the durface-normal direction, z. Let it be there, if referee complains then we will remove it.}}
Here, $\tau(z)$ is the positive KS
kinetic-energy density,
$\tau^W(z)=|dn(z)/dz|^2/[8n(z)]$ is the
von Weizs\"acker kinetic-energy density, and
$\tau^{\mathrm{unif}}(z)=
3[k_F^{\mathrm{loc}}(z)]^2n(z)/10$
is the corresponding uniform-electron-gas
kinetic-energy density.

To isolate the contribution associated with
the surface, we subtract the uniform bulk
exchange hole from the exchange hole of the
inhomogeneous slab. The resulting dimensionless
radial quantity is
\begin{equation}
\begin{split}
B(u)=k_F\int_{-\infty}^{\infty} dz\,\rho(z)
\Bigl[
&\rho(z)
J\bigl(p(z),\alpha(z),u\rho(z)^{1/3}\bigr)\\
&-J_b(u)
\Bigr],
\end{split}
\label{eq:Bslab}
\end{equation}
where $J(p,\alpha,y)$ is the dimensionless
spherically averaged exchange-hole shape function
and $J_b(u)=J(0,1,u)$ represents its uniform-gas
limit. In the uniform electron gas, the density
gradient vanishes ($p=0$) and $\alpha=1$.
The integration over $z$ includes the slab and
the surrounding vacuum. Therefore, $B(u)$
describes the bulk-subtracted exchange-hole
distribution as a function of the scaled
electron--electron separation $u$.

The corresponding wave-vector-resolved exchange
surface energy is obtained by a radial Fourier
transformation,
\begin{equation}
\gamma_x(k)=
\frac{4\bar n^2}{k_F^3}
\int_0^\infty du\,u^2B(u)
\operatorname{sinc}
\left(\frac{ku}{k_F}\right),
\label{eq:gammax}
\end{equation}
where $k$ is the magnitude of the Fourier
wave vector and
$\operatorname{sinc}(x)=\sin(x)/x$, with
$\operatorname{sinc}(0)=1$. In Fig.~\ref{fig:wv},
the spectrum is displayed against the
dimensionless wave vector $k/(2k_F)$.
The plotted $\gamma_x(k)$ is defined per
unit of this normalized wave vector and
is expressed in $\mathrm{erg/cm^2}$.

The total exchange surface energy per unit
area is obtained by integrating the spectrum
over the normalized wave vector,
\begin{equation}
\sigma_x=
\int_0^\infty
d\left(\frac{k}{2k_F}\right)
\gamma_x(k).
\label{eq:sigma}
\end{equation}
The factor in Eq.~\eqref{eq:gammax} assigns half of the bulk-subtracted slab energy to each of its two equivalent surfaces. Here, $\sigma_x$ is the integrated exchange
surface energy, whereas $\gamma_x(k)$ gives
its distribution among different wave-vector
components. Small values of $k/(2k_F)$
correspond to long-wavelength contributions,
while larger values probe shorter spatial
scales. The integrated surface energy is
therefore determined by the area under
the wave-vector spectrum.

The exchange-hole particle sum rule provides
an additional consistency condition. Since
both the slab and uniform-gas exchange holes
are normalized, the bulk-subtracted quantity
satisfies
\begin{equation}
\int_0^\infty du\,u^2B(u)=0.
\end{equation}
Consequently, Eq.~(\ref{eq:gammax}) gives
$\gamma_x(0)=0$ in the complete-integration
limit. At finite wave vectors, however,
$\gamma_x(k)$ may take either positive or
negative values because the Fourier-transform
kernel oscillates, even when the real-space
exchange hole remains nonpositive.

Figure~\ref{fig:wv} compares the wave-vector-resolved exchange
surface energy obtained from the present mSCAN exchange hole
with the exact-exchange and other semilocal results. The
calculated surface exchange energy,
$\sigma_x^{\ms}=2319\,\mathrm{erg\,cm^{-2}}$, is close to
the exact orbital-exchange value of
$2348\,\mathrm{erg\,cm^{-2}}$~\cite{jana2022}.
The mSCAN spectrum follows the overall behavior of the
exact-exchange curve, with its dominant contribution
arising from intermediate wave vectors. This suggests
that the model captures the principal redistribution
of the exchange hole associated with the formation of
the surface. However, noticeable differences appear
near $q=k/(2k_F)\simeq 1$, where the exact spectrum
exhibits a more pronounced change in its wave-vector
dependence. Since the wave-vector spectrum is obtained
from the Fourier transform of the bulk-subtracted
exchange hole, these differences indicate that the
model does not fully reproduce the spatial distribution
of exchange at all length scales. Furthermore, positive
and negative spectral contributions can partially cancel
upon integration, leading to an accurate total surface
energy even when the wave-vector dependence differs
from the exact result. The comparison with LDA, TPSS,
and MGGAC~\cite{jana2022} further illustrates the
sensitivity of the exchange surface energy to the
underlying exchange-hole description. Thus, the present
mSCAN hole provides a reasonable description of the
integrated exchange surface energy, although its
wave-vector distribution retains deviations from the
exact orbital-exchange behavior.

\subsection{Analytic short-range moment and its gradient response}\label{sec:screen}
The analytic short-range exchange can be obtained from the separation of the Coulomb interaction according to $1/u=\operatorname{erfc}(\omega u)/u+\operatorname{erf}(\omega u)/u$. Defining $\nu=\omega/k_F$, the same hole becomes
\begin{subequations}\label{eq:split}
\begin{align}
\Fx^{\ms,\Sr}(p,\alpha,\nu)&=-\frac89\int_0^\infty dy\,yJ(p,\alpha,y)\operatorname{erfc}(\nu y),\\
\Fx^{\ms,\Lr}&=\Fx^{\ms}-\Fx^{\ms,\Sr}.
\end{align}
\label{Eq:scan-sr}
\end{subequations}
Here $\nu$ is density dependent at fixed global $\omega$. Analytic integration of Eq.~\eqref{eq:hole}, with $\chi=\nu/(\lambda+\nu^2)^{1/2}$, yields
\begin{equation}
\begin{aligned}
\Fx^{\ms,\Sr}={}&A-\frac{4B}{9\lambda}(1-\chi)\\
&-\frac{4U}{9\lambda^2}\left(1-\frac{3\chi}{2}+\frac{\chi^3}{2}\right)\\
&-\frac{8V}{9\lambda^3}\left(1-\frac{15\chi}{8}+\frac{5\chi^3}{4}-\frac{3\chi^5}{8}\right)\\
&+2\nu\bigl(\sqrt{H+\nu^2}-\sqrt{\eta+\nu^2}\bigr)\\
&+2H\ln\frac{\nu+\sqrt{H+\nu^2}}{\nu+\sqrt{\lambda+\nu^2}}\\
&-2\eta\ln\frac{\nu+\sqrt{\eta+\nu^2}}{\nu+\sqrt{\lambda+\nu^2}}.
\end{aligned}\label{eq:srclosed}
\end{equation}
%The continuous $H=0$ convention is understood. 
In Eq.~\ref{eq:srclosed}, the limits $\nu\to0$ and $\nu\to\infty$ give $\Fx^{\ms,\Sr}=\Fx^{\ms}$ and zero, respectively. %We independently checked Eq.~\eqref{eq:srclosed} against direct numerical integration of Eq.~\eqref{eq:split} at $\nu=0.4$; for the representative points reported in the Supplemental Material, differences are below $10^{-8}$.

\subsection{Screened hybrid functional}\label{sec:hybrid}
The resulting hybrid functional is
\begin{align}
E_{xc}^{\rm HSE-\ms}
={}&E_x^{\ms}+E_c^{\rm modSCAN}\nonumber\\
&+a_x(E_x^{\rm HF,\Sr}-E_x^{\ms,\Sr})~,\label{eq:hybrid}
\end{align}
where the short-range (SR) orbital exchange is
\begin{equation}
E_x^{\rm HF,\Sr}=-\frac12\sum_\sigma\iint d^3r\,d^3r'\,|\gamma_\sigma(\mathbf r,\mathbf r')|^2
\frac{\operatorname{erfc}(\omega|\mathbf r-\mathbf r'|)}{|\mathbf r-\mathbf r'|},
\end{equation}
and $\gamma_\sigma(\mathbf r,\mathbf r')=\sum_i f_{i\sigma}\phi_{i\sigma}(\mathbf r)\phi^*_{i\sigma}(\mathbf r')$. Exchange spin scaling is applied separately to the two semilocal components,
\begin{equation}
 E_x^X[n_\uparrow,n_\downarrow]
 =\tfrac12 E_x^X[2n_\uparrow]+\tfrac12 E_x^X[2n_\downarrow],
 \label{eq:spin-scaling}
\end{equation}
Here $X$ denotes full-range or short-range mSCAN exchange; gradients and kinetic-energy densities are scaled consistently within each spin channel. %The present benchmarks use collinear spin states.

We use $a_x=0.10$ and $\omega=0.11\,\mathrm{bohr}^{-1}$, the parameters used in the earlier MGGAC screened hybrid~\cite{jana2022}. The screening parameter is the conventional HSE06 value~\cite{krukau2006}. Section~S1 of the Supplemental Information (SI)~\cite{supplement} tests the exchange mixing by weakly modulating a uniform electron gas. Figure~S1(a) of the SI shows that the correlation gradient contribution offsets much of the finite-wave-vector exchange response. For example, at $r_s=2.07$ and $Q=2k_F$, including this contribution lowers the calculated local-field factor from 1.8110 to 0.9395, close to the Corradini reference~\cite{corradini1998} of 0.9503. Figure~S1(b) shows how the response varies with wave vector as the mixing fraction changes. In Table~S1, at $r_s=3$ and $Q=2k_F$, replacing 10\% of the semilocal short-range exchange with Fock exchange gives 0.9675, compared with 0.9985 for zero mixing and 0.9209 for 25\% mixing (reference 0.9820). This density- and wave-vector-dependent balance motivates a modest Fock fraction. The fixed-orbital energy-curvature test leaves the optimal fraction undetermined.
In addition, the SCAN correlation interpolation parameter $c_{1c}$ changes from 0.64 to 1.60. This defines a modified correlation functional, $E_c^{\mathrm{modSCAN}}$. These fixed choices are part of the present approximation; neither exchange-hole normalization nor recovery of the exchange-energy moment determines them uniquely.

To specify
precisely which part is modified, we write the correlation
energy per electron as
\begin{equation}
 \epsilon_c^{\mathrm{modSCAN}}
 =\epsilon_c^1+f_c^{\mathrm{int}}(\alpha;c_{1c})
 (\epsilon_c^0-\epsilon_c^1),
 \label{eq:modSCAN-correlation}
\end{equation}
where $\epsilon_c^0$ and $\epsilon_c^1$ are the single-orbital
($\alpha=0$) and slowly varying ($\alpha\simeq1$) correlation
branches of SCAN, respectively. The quantity
$f_c^{\mathrm{int}}$ denotes the correlation interpolation
factor of the SCAN functional~\cite{scan2015}:
\begin{equation}
 f_c^{\mathrm{int}}(\alpha;c_{1c})=
 \begin{cases}
 \exp[-c_{1c}\alpha/(1-\alpha)], & 0\leq\alpha<1,\\
 0, & \alpha=1,\\
 -d_c\exp[c_{2c}/(1-\alpha)], & \alpha>1.
 \end{cases}
 \label{eq:modSCAN-interpolation}
\end{equation}
Here $d_c=0.7$ and $c_{2c}=1.5$ retain their SCAN values
\cite{scan2015}. Increasing $c_{1c}$ reduces the weight of the single-orbital correlation branch for $0<\alpha<1$, shifting the interpolation toward $\epsilon_c^1$ in regions intermediate between single-orbital and slowly varying densities. Both endpoint values and the $\alpha>1$ branch retain their SCAN forms. The interpolation factor and its derivatives vanish at $\alpha=1$, so the correlation gradient contribution that compensates exchange in the quadratic uniform-gas response is independent of $c_{1c}$ (Sec.~S1 of the SI~\cite{supplement}). The value 1.60 therefore sets the interpolation in inhomogeneous regions and remains undetermined by this response test.

%At the semilocal level, mSCAN exchange is fully specified by its enhancement factor and does not require an explicit hole. Its role here is the energy target for range separation. Consequently, a semilocal comparison alone cannot distinguish holes having the same full exchange moment. Conversely, the present hybrid benchmarks cannot isolate the benefit of the hole shape from the changes in exchange target, Fock fraction, and correlation. Matched-parameter controls would be needed for that attribution.

\section{Results and discussion}

\subsection{Computational details and reference sets}

Solid-state calculations are performed using VASP 6.2.0 with projector-augmented-wave (PAW) pseudopotentials~\cite{blochl1994,kresse1996cms,kresse1996,kresse1999}. A plane-wave kinetic-energy cutoff of 550 eV and a $\Gamma$-centered $12\times12\times12$ $k$-point mesh are employed~\cite{monkhorst1976}. The calculations are converged to approximately 0.01 eV with respect to the relevant numerical settings. For cohesive-energy calculations, isolated atomic reference systems are treated in orthorhombic supercells with dimensions of $15\times16\times17\,\mathrm{\mathring{A}^3}$. Bulk moduli are obtained by fitting the calculated energy--volume data to a fourth-order Birch--Murnaghan equation of state~\cite{birch1947}. Spin--orbit coupling (SOC) is not included, and no empirical SOC corrections are applied to the calculated band gaps.

The solid-state structural benchmark comprises 46 cubic solids, consisting of 10 simple or $sp$ metals, 14 transition metals, 15 semiconductors, and seven insulators. The electronic band-gap benchmark includes 76 solids, among which six are magnetic oxides. Experimental fundamental band gaps are taken from the compilation reported in Ref.~\cite{tran2017}. The detailed results for lattice constants, bulk moduli, cohesive energies, and band gaps are provided in the SI~\cite{supplement}. The reference data for the solid-state benchmarks are adopted from established compilations and previous benchmark studies~\cite{haas2009,haas2009erratum,schimka2011,csonka2009,Kaplan2022,Janthon2014,Zhang2018,peverati2012,tao2017}. 

Water-hexamer calculations use Q-Chem~\cite{qchem2021} and the aug-cc-pVTZ basis set~\cite{unc-aug-cc-pvtz}.

\subsection{Lattice constants, bulk moduli, and cohesive energies}

\begin{figure*}%[t]
\centering
\includegraphics[width=0.98\textwidth]{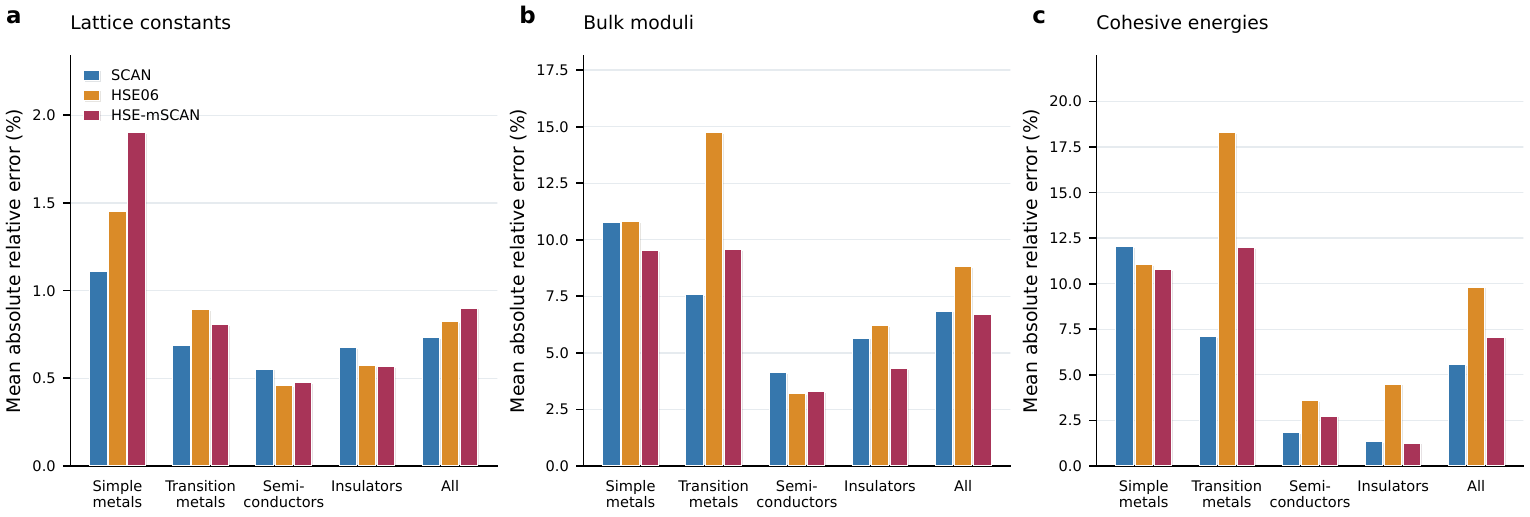}
\caption{Mean absolute relative errors (MAREs) of SCAN, HSE06, and HSE-mSCAN for (a) lattice constants, (b) bulk moduli, and (c) cohesive energies across different classes of solids. The results are shown separately for simple metals, transition metals, semiconductors, and insulators, together with the overall statistics for the complete benchmark set.}
\label{fig:solid-errors}
\end{figure*}

Table~\ref{tab:combined-stats} summarizes errors for the 46 cubic solids, and Fig.~\ref{fig:solid-errors} resolves the relative errors by material class. SCAN gives the smallest lattice-constant MAE (0.035~\AA), followed by HSE06 (0.039~\AA) and HSE-mSCAN (0.044~\AA). HSE-mSCAN reproduces the Cu lattice constant to within 0.001~\AA, but increases the overestimate for Cs from 0.200 to 0.278~\AA.
The bulk modulus probes the curvature of the energy--volume curve at equilibrium. HSE-mSCAN reduces the bulk-modulus MAE from 11.163 GPa with HSE06 to 8.218 GPa, close to the SCAN value of 7.467 GPa. Ni illustrates the distinction between curvature and equilibrium volume. HSE-mSCAN reduces the absolute bulk-modulus error from 63.8 to 13.0 GPa, although HSE06 gives the more accurate lattice constant.
Cohesive energies depend on the energy difference between isolated atoms and the crystal. HSE06 underbinds by 0.351 eV/atom on average. HSE-mSCAN reduces this bias to 0.222 eV/atom and the MAE from 0.396 to 0.273 eV/atom. The improvement is pronounced for Cu and Ta, while underbinding increases for Cs. SCAN gives the smallest overall MAE (0.198 eV/atom).
The change in large-gradient exchange affects atomic density tails and low-density regions between atoms in the crystal. Their exchange contributions are weighted by the local density. Because HSE-mSCAN also changes correlation and the fraction of short-range Fock exchange relative to HSE06, the benchmark cannot isolate the contribution of the exchange hole to these trends.

\begin{table*}[t]
\centering
\small
\setlength{\tabcolsep}{12pt}
\caption{Statistical errors for lattice constants, bulk moduli, cohesive energies, and band gaps. The first three properties use 46 cubic solids; band gaps use 76 solids. ME, MAE, and STDE are in the units indicated; MRE, MARE, and STDRE are percentages. Bold indicates the smallest absolute value for ME and MRE, and the smallest value for the other metrics. Statistics are recomputed from the numerical values tabulated in the Supplemental Material; standard deviations use the population convention.}
\label{tab:combined-stats}
\begin{tabular}{lrrrrrr}
\toprule
\multicolumn{7}{c}{Lattice constants (\AA)} \\
\midrule
Method & ME & MAE & STDE & MRE (\%) & MARE (\%) & STDRE (\%) \\
\midrule
SCAN      & \textbf{0.015} & \textbf{0.035} & 0.048 & \textbf{0.196} & \textbf{0.734} & 0.974 \\
HSE06     & 0.032 & 0.039 & \textbf{0.046} & 0.616 & 0.823 & \textbf{0.963} \\
HSE-mSCAN & 0.025 & 0.044 & 0.067 & 0.403 & 0.899 & 1.303 \\
\midrule
\multicolumn{7}{c}{Bulk moduli (GPa)} \\
\midrule
Method & ME & MAE & STDE & MRE (\%) & MARE (\%) & STDRE (\%) \\
\midrule
SCAN      & 2.271 & \textbf{7.467} & \textbf{12.301} & -1.316 & 6.845 & \textbf{9.044} \\
HSE06     & \textbf{-0.381} & 11.163 & 18.980 & -3.692 & 8.823 & 11.706 \\
HSE-mSCAN & 2.224 & 8.218 & 13.223 & \textbf{-0.406} & \textbf{6.705} & 9.297 \\
\midrule
\multicolumn{7}{c}{Cohesive energies (eV/atom)} \\
\midrule
Method & ME & MAE & STDE & MRE (\%) & MARE (\%) & STDRE (\%) \\
\midrule
SCAN      & \textbf{0.013} & \textbf{0.198} & \textbf{0.315} & \textbf{0.625} & \textbf{5.566} & \textbf{9.702} \\
HSE06     & -0.351 & 0.396 & 0.483 & -7.978 & 9.809 & 10.807 \\
HSE-mSCAN & -0.222 & 0.273 & 0.430 & -4.947 & 7.072 & 9.847 \\
\midrule
\multicolumn{7}{c}{Band gaps (eV)} \\
\midrule
Method & ME & MAE & STDE & MRE (\%) & MARE (\%) & STDRE (\%) \\
\midrule
SCAN      & -1.541 & 1.541 & 1.362 & -40.691 & 40.701 & \textbf{22.145} \\
HSE06     & \textbf{-0.695} & \textbf{0.838} & \textbf{1.201} & \textbf{-6.868} & \textbf{18.525} & 24.367 \\
HSE-mSCAN & -0.860 & 0.932 & 1.210 & -12.235 & 20.943 & 24.968 \\
\bottomrule
\end{tabular}
\end{table*}

\subsection{Band gaps of semiconductors and insulators}

The performance of screened hybrid functionals and generalized Kohn--Sham (GKS) approaches for predicting electronic band gaps has been extensively investigated for solids~\cite{heyd2005,paier2006,fuchs2007}, while semilocal model potentials provide an alternative approach to calculating one-electron spectra~\cite{tran2009}. Here, we compare SCAN, HSE06, and HSE-mSCAN for the same set of 76 solids, using the experimental fundamental band gaps compiled in Ref.~\cite{tran2017} as reference values. Neither spin--orbit coupling (SOC) nor empirical a posteriori SOC corrections are included in the calculated band gaps. The corresponding reference structural space groups are provided in the Supplemental Material~\cite{supplement}.

The overall mean absolute errors (MAEs) are 1.541, 0.838, and 0.932 eV for SCAN, HSE06, and HSE-mSCAN, respectively. Thus, HSE-mSCAN substantially improves upon SCAN but retains a slightly larger MAE than HSE06. All three functionals underestimate the experimental fundamental gaps on average, with mean signed errors of $-1.541$, $-0.695$, and $-0.860$ eV, respectively. The largest underestimations with HSE-mSCAN occur for Ne ($-7.35$ eV), Ar ($-3.74$ eV), NaF ($-3.37$ eV), and KF ($-3.18$ eV). These systems are retained in the full-set statistics and parity plot. Relative errors are also reported to provide a complementary measure of the deviations across the benchmark set. However, they become particularly sensitive to small absolute differences for narrow-gap materials, such as InSb. Finally, the absence of SOC is an important consideration when interpreting the results for compounds containing heavy elements, particularly those with narrow gaps. No empirical SOC correction is applied, ensuring that the reported comparison reflects the calculated results without additional adjustments.

\begin{figure*}%[t]
\centering
\includegraphics[width=0.86\linewidth]{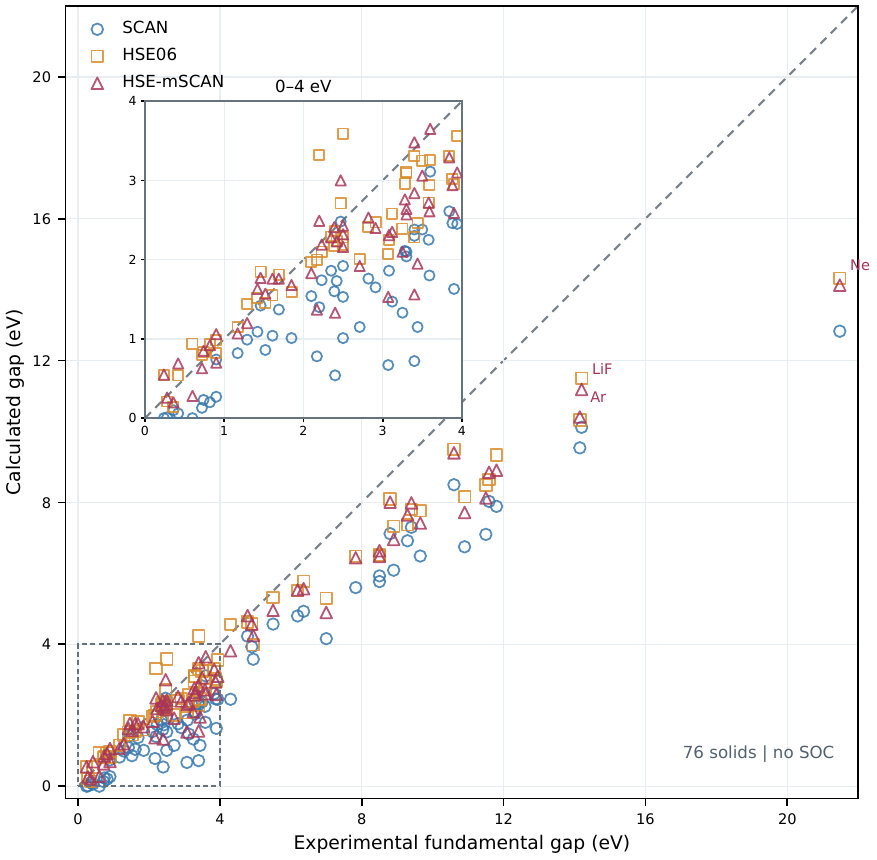}
\caption{Calculated versus experimental fundamental band gaps for all 76 solids. The dashed line denotes equality. The calculation data are used without fitted shifts or SOC corrections; experimental references are from Tran and Blaha~\cite{tran2017}. The inset resolves the 0--4 eV range. The broad axis ranges retain the rare-gas and wide-gap entries.}
\label{fig:gap-parity}
\end{figure*}

\subsection{Transition-metal oxides and local spin moments}

Table~\ref{tab:tmo-gaps-moments} compares calculated band gaps and local spin moments with experiment for the six magnetic oxides in the 76-solid benchmark.

For this subset, the gap MAE decreases from 1.548 eV with SCAN and 0.715 eV with HSE06 to 0.550 eV with HSE-mSCAN. HSE-mSCAN gives gaps closer to the selected experimental values than HSE06 for CoO, Cr$_2$O$_3$, and Fe$_2$O$_3$, whereas HSE06 performs better for FeO, MnO, and NiO. Thus, the lower MAE of HSE-mSCAN for the magnetic oxides does not translate into a lower MAE for the full set of 76 solids.

\begin{table*}%[t]
\centering
\small
\setlength{\tabcolsep}{4.5pt}
\caption{Calculated band gaps and local magnetic moments of six magnetic oxides. Experimental band gaps are taken from Ref.~\cite{tran2017}, and experimental ordered moments are from neutron-diffraction measurements~\cite{KhanErickson1970,Roth1958,Brown2002,Hill2008Hematite}. The calculated moments are site-averaged absolute PAW-projected spin moments, in $\mu_{\mathrm B}$ per transition-metal atom. These projections are not strictly equivalent to the experimental ordered moments and may differ because of orbital contributions and the spatial definition of the calculated moments. Bold indicates the calculated band gap closest to experiment in each row.}
\label{tab:tmo-gaps-moments}
\begin{tabular}{lccrrrrrr}
\toprule
& \multicolumn{2}{c}{Experiment}
& \multicolumn{2}{c}{SCAN}
& \multicolumn{2}{c}{HSE06}
& \multicolumn{2}{c}{HSE-mSCAN} \\
\cmidrule(lr){2-3}\cmidrule(lr){4-5}\cmidrule(lr){6-7}\cmidrule(lr){8-9}
Material & Gap (eV) & Moment ($\mu_{\mathrm B}$)
& Gap (eV) & Moment ($\mu_{\mathrm B}$)
& Gap (eV) & Moment ($\mu_{\mathrm B}$)
& Gap (eV) & Moment ($\mu_{\mathrm B}$) \\
\midrule
CoO             & 2.50 & 3.35 & 1.01 & 2.588 & 3.59 & 2.668 & \textbf{2.43} & 2.669 \\
FeO             & 2.40 & 3.32 & 0.54 & 3.546 & \textbf{2.36} & 3.630 & 1.33 & 3.616 \\
MnO             & 3.90 & 4.58 & 1.63 & 4.458 & \textbf{2.95} & 4.523 & 2.59 & 4.530 \\
NiO             & 4.30 & 1.90 & 2.45 & 1.572 & \textbf{4.56} & 1.657 & 3.82 & 1.654 \\
Cr$_2$O$_3$     & 3.40 & 2.48 & 2.38 & 2.729 & 4.23 & 2.828 & \textbf{3.48} & 2.788 \\
$\alpha$-Fe$_2$O$_3$ & 2.20 & 4.22 & 1.40 & 3.927 & 3.32 & 4.124 & \textbf{2.49} & 4.107 \\
\bottomrule
\end{tabular}
\end{table*}

The calculated local moments are obtained from the VASP magnetization projections by averaging the absolute site-resolved spin moments over the transition-metal atoms. Although the magnetic sublattices carry opposite spin orientations, their absolute moments remain finite, while the total cell moment is approximately zero. HSE-mSCAN gives moments of 2.669, 3.616, 4.530, and 1.654 $\mu_{\rm B}$ per transition-metal atom for CoO, FeO, MnO, and NiO, respectively, and 2.788 and 4.107 $\mu_{\rm B}$ for Cr$_2$O$_3$ and Fe$_2$O$_3$. The differences between HSE-mSCAN and HSE06 are small, with a maximum absolute difference of 0.040 $\mu_{\rm B}$ across these six oxides.

A direct quantitative comparison with experiment requires some care because neutron-diffraction ordered moments and PAW on-site spin projections are different observables. Experimental values can depend on temperature, composition, covalency, and the refinement model, and may include orbital contributions~\cite{Roth1958,KhanErickson1970,Brown2002,Hill2008Hematite}. Recent refinements for MnO and NiO, together with other oxide calculations, also illustrate the variation in reported reference moments~\cite{Pomjakushin2024MnONiO,Liu2020Oxides,Naveas2023,Dannegger2023Hematite}. We therefore use the calculated moments to characterize the local spin polarization rather than to establish an aggregate ranking of moment errors. Since SOC is not included in the present calculations, orbital magnetism and weak spin canting are not described.

\subsection{Phase stability of \texorpdfstring{MnO$_2$}{MnO2} polymorphs}

\begin{figure*}%[t]
\centering
\includegraphics[width=0.82\textwidth]{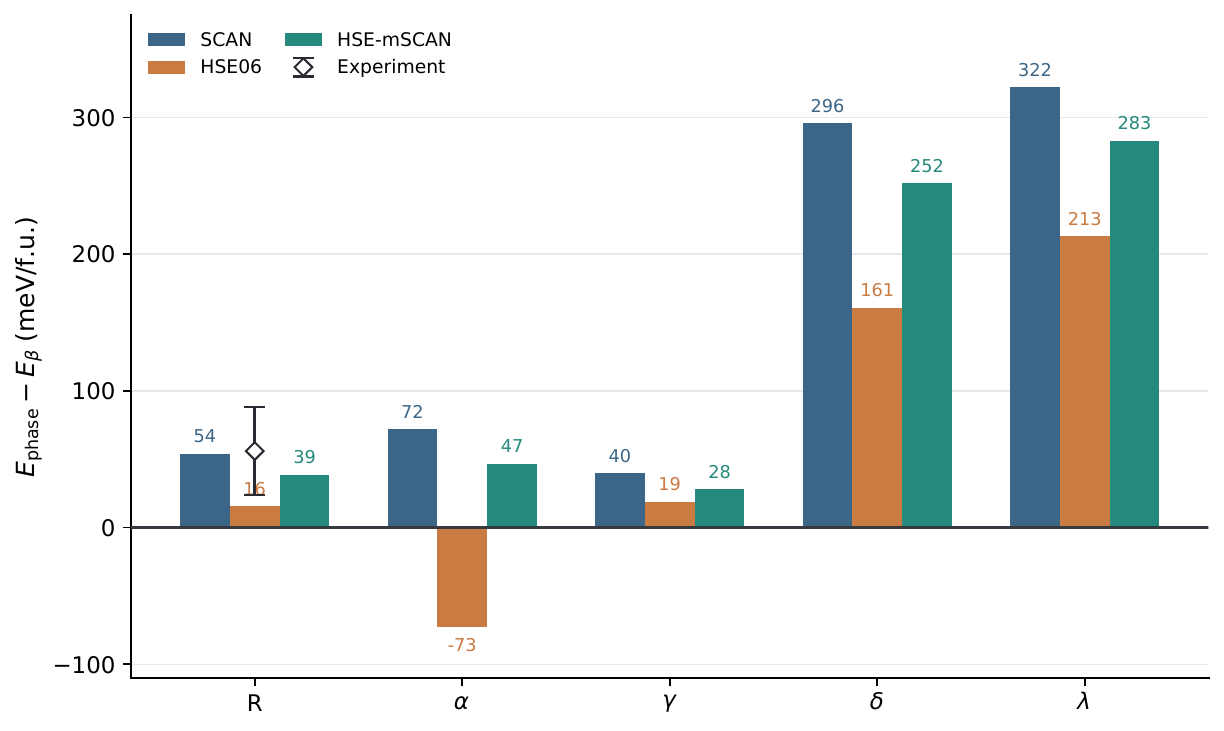}
\caption{Electronic energies of MnO$_2$ polymorphs relative to the $\beta$ phase, in meV per formula unit, calculated at fixed geometries. SCAN and HSE06 values are digitized from Ref.~\cite{Kitchaev2016}, while HSE-mSCAN values are obtained from the present calculations for the same selected geometries and magnetic configurations. The experimental R--$\beta$ enthalpy difference is $56\pm32$ meV/f.u.; the experimental marker is shown only for ramsdellite (R). No experimental energy differences are available for the other polymorphs. The estimated uncertainty associated with digitizing the literature data is approximately 2 meV/f.u. The SCAN R--$\beta$ energy difference of 54.2 meV/f.u. is taken directly from the text of Ref.~\cite{Kitchaev2016}.}
\label{fig:phase-stability}
\end{figure*}

Figure~\ref{fig:phase-stability} compares six MnO$_2$ polymorphs at fixed geometries and selected magnetic configurations, using the SCAN and HSE06 results of Ref.~\cite{Kitchaev2016}. We report electronic energies per formula unit relative to the $\beta$ phase,
\begin{equation}
 \Delta E_i=e_i-e_\beta,\qquad
 e_i=\frac{E_i^{\rm cell}}{N_i},
\end{equation}
where $N_i$ is the number of MnO$_2$ formula units in the calculation cell. Positive values indicate that phase $i$ is higher in energy than $\beta$.

HSE-mSCAN gives the phase ordering
\begin{equation}
 \beta < \gamma < {\rm R} < \alpha < \delta < \lambda,
\end{equation}
with relative energies of $28.34$, $38.81$, $46.86$, $251.86$, and $283.18$ meV/MnO$_2$ for $\gamma$, ramsdellite (R), $\alpha$, $\delta$, and $\lambda$, respectively [Fig.~\ref{fig:phase-stability}]. Thus, HSE-mSCAN predicts $\beta$ to be the lowest-energy phase among the structures and magnetic configurations considered. In contrast, HSE06 places $\alpha$ approximately $73$ meV/MnO$_2$ below $\beta$, reversing their relative stability, whereas SCAN correctly favors $\beta$~\cite{Kitchaev2016}. HSE-mSCAN therefore reproduces the SCAN phase ordering and corrects the HSE06 prediction. The relatively small energy differences among the tunnel and intergrowth structures contrast with the substantially higher energies of the layered $\delta$ and spinel $\lambda$ phases.

The experimental ramsdellite--$\beta$ enthalpy difference is $56\pm32$ meV per formula unit~\cite{Kitchaev2016}. The electronic energy differences from HSE-mSCAN (38.81 meV) and SCAN (54.2 meV) both fall within this interval. Experimental energy differences for the other phases are unavailable in the cited study.

\subsection{Relative stability of water-hexamer isomers}

The water hexamer provides a test of how well a functional describes the relative stabilization of distinct hydrogen-bond networks~\cite{santra2008,bates2009,gillan2016water}. The CCSD(T)/CBS reference places the prism isomer lowest in energy, followed by cage, book-1, book-2, bag, cyclic-chair, cyclic-boat-1, and cyclic-boat-2~\cite{Jana2020}. Since all isomers contain the same six water molecules, their relative energies probe the balance of interactions among different hydrogen-bond arrangements. We therefore compare
\begin{equation}
 \Delta E_i=E_i-E_{\rm prism},\qquad
 \delta_i=\Delta E_i-\Delta E_i^{\rm ref}.
\end{equation}

\begin{figure*}%[t]
\centering
\includegraphics[width=0.86\linewidth]{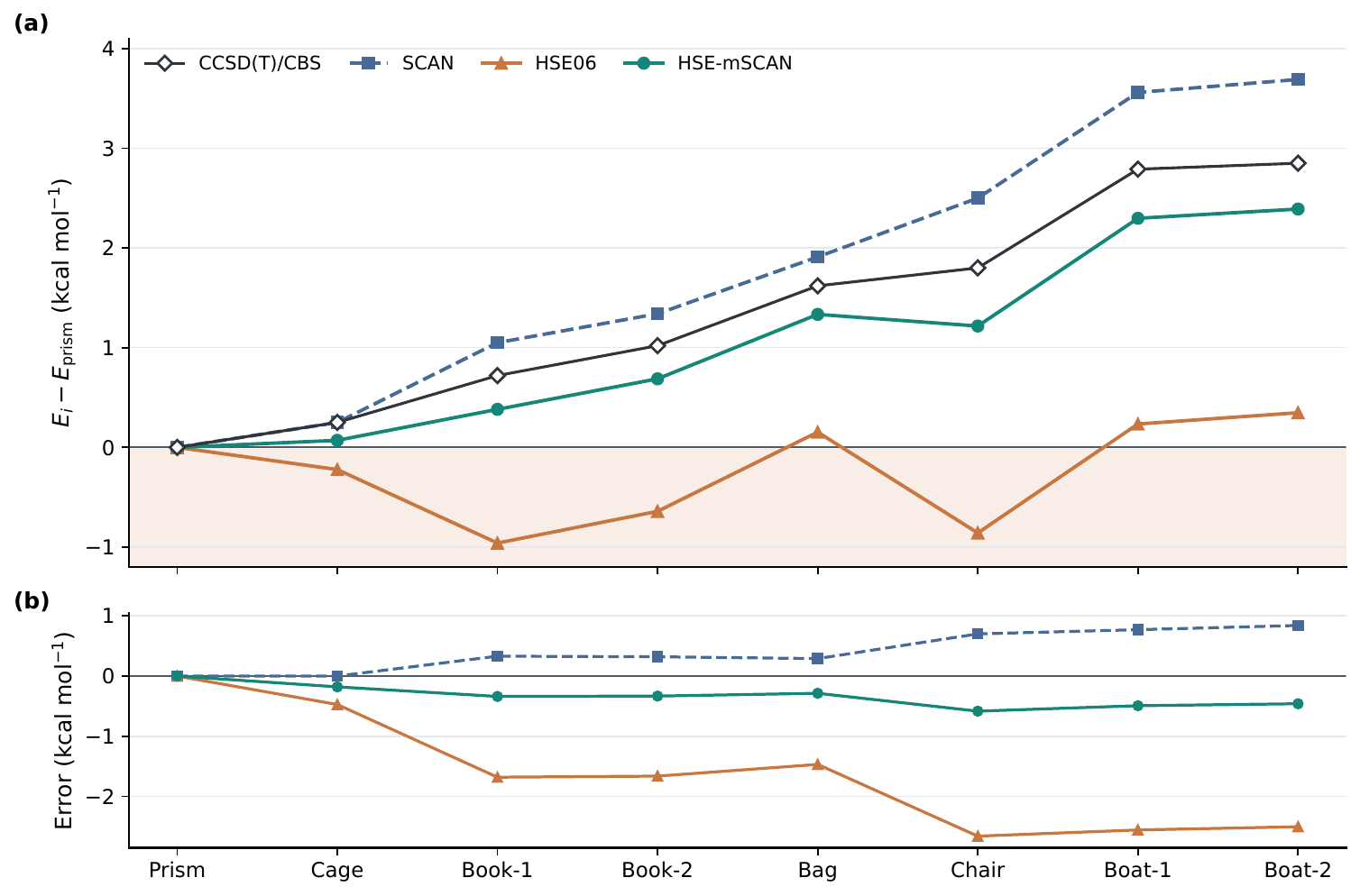}
\caption{Relative electronic energies of eight water-hexamer isomers. \textbf{a}, Energies relative to the prism isomer, $\Delta E_i=E_i-E_{\rm prism}$, in kcal/mol per hexamer. The shaded region indicates isomers lying below the prism. \textbf{b}, Signed errors, $\Delta E_i-\Delta E_i^{\rm ref}$, in kcal/mol. The CCSD(T)/CBS reference energies and SCAN results are taken from the water-hexamer benchmark reported by ref.~\cite{santra2008} and the corresponding data in Table~3 of ref.~\cite{Jana2020}. HSE06 and HSE-mSCAN results are from the present calculations. Chair and boat labels denote cyclic isomers. Lines connect the discrete isomers as visual guides and do not represent transformation pathways. The prism energy and its error are zero by definition and are excluded from the error statistics.}
\label{fig:water-hexamer}
\end{figure*}

As shown in Fig.~\ref{fig:water-hexamer}, HSE06 incorrectly predicts four isomers---cage, book-1, book-2, and cyclic-chair---to lie below the prism. Its lowest-energy structure is book-1, which is $0.960$ kcal/mol below the prism, whereas the reference places it $0.720$ kcal/mol above. This reversal indicates an incorrect balance of relative stabilization among the isomers. HSE06 also substantially underestimates the energy separation of the cyclic-boat isomers from the prism, predicting $0.234$ and $0.348$ kcal/mol compared with the reference values of $2.790$ and $2.850$ kcal/mol.

In contrast, HSE-mSCAN retains the prism as the lowest-energy isomer and places the cage only $0.071$ kcal/mol above it. The MAE over the seven non-prism energy differences decreases from $1.857$ kcal/mol for HSE06 to $0.382$ kcal/mol for HSE-mSCAN. The corresponding MAE for the literature SCAN results is $0.464$ kcal/mol, calculated using the values in Table~3 of Ref.~\cite{Jana2020}. Although SCAN reproduces the reference ordering, it generally overestimates the energy separations. HSE-mSCAN, on the other hand, underestimates all seven energy differences and reverses the relative ordering of bag and cyclic-chair, placing cyclic-chair $0.117$ kcal/mol below bag instead of $0.180$ kcal/mol above it. Thus, HSE-mSCAN substantially improves the relative energies and correctly identifies the prism minimum, although some errors in the detailed isomer ordering remain.

This improvement is specific to the relative energies of the eight isomers considered here. It does not establish the accuracy of absolute binding energies or the description of condensed-phase water, for which dispersion and many-body interactions also matter~\cite{dion2004,grimme2010,vydrov2010,klimes2011,peng2016rvv10,jana2020dispersion}. Moreover, the comparison involves changes in the semilocal exchange, Fock-exchange fraction, and correlation. The relative-energy results alone therefore do not allow the improvement to be attributed specifically to the exchange-hole construction.

%The SCAN results are literature single-point calculations performed using MP2/aug-cc-pVTZ geometries and the aug-cc-pVTZ basis set~\cite{Jana2020}. The supplied HSE06 and HSE-mSCAN energy files do not specify the geometries or basis sets used. These details should be documented alongside the molecular computational protocol to establish the extent to which the comparison is consistent across the calculations. No zero-point or thermal corrections are included.

\section{Conclusions}\label{sec:conclusion}

We have constructed an analytic exchange-hole model for modified SCAN exchange with explicit dependence on the reduced density gradient and the orbital kinetic-energy indicator. The on-top value, particle sum rule, and exchange-energy moment are imposed algebraically, while numerical tests support nonpositivity over the sampled parameter domain. Applying the same restricted construction to the original SCAN exchange target produces positive regions in the exchange hole. This result identifies a limitation of the present ansatz rather than establishing a general impossibility of constructing nonpositive SCAN exchange holes. The omission of the original SCAN damping factor also modifies the parent semilocal functional, including its dimensional-crossover behavior.

The analytic screened moment provides the semilocal SR exchange contribution required for HSE-mSCAN. Its dependence on the reduced gradient persists even in regions where the full exchange enhancement factor is constant, illustrating that the exchange-energy density alone does not uniquely determine the range-separated exchange. The calculated jellium-slab exchange energy agrees closely with the orbital-exchange reference, although differences remain in the wave-vector-resolved spectrum. The uniform-electron-gas curvature analysis reveals XC compensation and a density-dependent sensitivity to Fock mixing. However, it does not uniquely determine the selected 10\% Fock-exchange fraction or the modified correlation contribution.

The molecular and solid-state benchmarks show that the fixed-parameter HSE-mSCAN functional gives property-dependent performance. Relative to HSE06, it reduces the MAEs for bulk moduli and cohesive energies to 8.218 GPa and 0.273 eV/atom, respectively, but increases those for lattice constants and the full band-gap set to 0.044~\AA\ and 0.932 eV, respectively. For the six magnetic oxides, HSE-mSCAN gives a lower band-gap MAE than both SCAN and HSE06. It also improves the relative energies of water-hexamer isomers, retaining the prism as the lowest-energy structure, although the complete reference ordering is not reproduced. For MnO$_2$, it corrects the HSE06 prediction of the relative stability of the $\alpha$ and $\beta$ phases. These results demonstrate improvements for selected properties and systems, rather than a uniform improvement across the benchmarks.

Overall, this work provides an analytic SCAN-derived exchange hole and a corresponding screened-exchange construction for a meta-GGA hybrid functional. An important direction for future work is to extend this exchange-hole construction to screened dielectric-dependent hybrid (DDH), long-range-corrected (LRC), and double-hybrid (DH) functionals. Its analytic description of the screened semilocal exchange contribution provides a useful starting point for developing range-separated exchange schemes in which the screening and long-range behavior can be treated consistently. Work along these directions is ongoing, and the resulting developments and applications will be reported in future studies.

\section*{Data availability}

All data supporting the findings of this study are available within the article and its Supporting Information~\cite{supplement}. Additional data, including the structural files used in the calculations, will be made available in an external repository upon publication of the manuscript.

\section*{Acknowledgments}

This work was supported by the Polish National Science Center (NCN) under
Grant No.\ 2021/42/E/ST4/00096

\bibliographystyle{apsrev4-2}\bibliography{references}
\end{document}

\title{
Supplemental Material: Exchange-hole model for modified SCAN: Real-space constraints and screened hybrid exchange
}

\author{Subrata Jana}
\email{subrata.niser@gmail.com, subrata.jana@umk.pl}
\affiliation{Institute of Physics, Faculty of Physics, Astronomy and Informatics, Nicolaus Copernicus University in Toru\'n, ul. Grudzi\k{a}dzka 5, 87-100 Toru\'n, Poland}
 \affiliation{Institute of Advanced Studies, Nicolaus Copernicus University in Toru\'{n}, ul. Wile\'{n}ska 4, 87-100 Toru\'{n}, Poland} 

\author{Szymon \'Smiga}
\affiliation{Institute of Physics, Faculty of Physics, Astronomy and Informatics, Nicolaus Copernicus University in Toru\'n, ul. Grudzi\k{a}dzka 5, 87-100 Toru\'n, Poland}
 \affiliation{Institute of Advanced Studies, Nicolaus Copernicus University in Toru\'{n}, ul. Wile\'{n}ska 4, 87-100 Toru\'{n}, Poland} 

\date{\today}

\maketitle

\renewcommand{\thefigure}{S\arabic{figure}}\renewcommand{\thetable}{S\arabic{table}}\renewcommand{\theequation}{S\arabic{equation}}

\renewcommand{\thesection}{S\arabic{section}}

\onecolumngrid

\section{Finite-wave-vector response and short-range exchange mixing}
%\szs{Is this subsection necessary? }
\label{sec:finiteQ-mixing}

The finite-wave-vector XC response provides an additional test of the balance between exchange and correlation in the HSE-mSCAN functional. Although the exchange-hole constraints determine the parent semilocal exchange energy and its short-range partition, they do not by themselves establish how the complete functional responds to spatial variations in the electron density. In this respect, the gradient contribution of the modified SCAN correlation plays an important role by compensating a substantial part of the finite-wave-vector exchange response. Introducing short-range Fock exchange modifies this balance in a wave-vector- and density-dependent manner. Comparison with the Corradini analytical parametrization~\cite{corradini1998} shows that a modest fraction of Fock exchange can limit deviations in the XC response, whereas increasing the Fock fraction does not necessarily improve the agreement. These observations provide a physical motivation for choosing $a_x=0.10$, although the present non-self-consistent energy-curvature analysis does not establish a unique optimal mixing fraction.

To quantify the effect of short-range Fock exchange on the XC response, we examine how the mixing fraction $a_x$ in Eq.(24) of the main text %Eq.~\eqref{eq:hybrid} 
influences the XC energy of a uniform electron gas subjected to a weak density modulation. Such an analysis of the uniform-gas XC kernel provides a useful means of assessing the response of hybrid functionals to density variations and has previously been employed to examine the choice of exchange mixing~\cite{moldabekov2023}. Specifically, for a spin-unpolarized electron gas, we consider
\begin{equation}
 n(z)=n_0+\delta n\cos(Qz),\qquad |\delta n|\ll n_0,
 \label{eq:finiteQ-modulation}
\end{equation}
where $n_0$ is the unperturbed density, $\delta n$ is the density
modulation amplitude, and $Q$ is the wave-vector magnitude along
$z$. The perturbed noninteracting orbitals are obtained in the
potential $v_s(z)=2V_Q\cos(Qz)$, where $V_Q$ is a small potential
amplitude; the mean density is held fixed. In atomic units,
$r_s=[3/(4\pi n_0)]^{1/3}$ is the Wigner--Seitz radius and
$k_F=(3\pi^2n_0)^{1/3}$ is the Fermi wave vector of the unperturbed
gas. The ratio $Q/k_F$ describes the modulation length scale:
$Q\ll k_F$ corresponds to a slowly varying density, whereas
$Q\simeq2k_F$ is associated with scattering across the Fermi
surface.

For an infinitesimal density modulation, the XC energy change
of the functional in Eq.(24) %Eq.~\eqref{eq:hybrid} 
is written as
\begin{equation}
 \frac{\Delta E_{xc}}{\mathcal V}
 =\frac{(\delta n)^2}{4}\,\mathcal K_{xc}(Q;a_x)
 +o\!\left((\delta n)^2\right),
 \label{eq:finiteQ-curvature}
\end{equation}
where $\mathcal V$ is the system volume and $\Delta E_{xc}$ is
measured relative to the unperturbed gas. The quantity
$\mathcal K_{xc}(Q;a_x)$ is the quadratic XC energy curvature
along the density and orbital response generated by $v_s$.
The factor $1/4$ follows from the quadratic expansion and the
spatial average of $\cos^2(Qz)$. 

For comparison with uniform-gas reference data,
we define its dimensionless local-field-factor representation,
\begin{equation}
 G(Q;a_x)
 =-\frac{Q^2}{4\pi}\,\mathcal K_{xc}(Q;a_x).
 \label{eq:finiteQ-Gcalc}
\end{equation}

Following the decomposition of the HSE-mSCAN energy in
Eq. (24) %~\eqref{eq:hybrid}  \szs{which one???}, 
the calculated curvature is
\begin{equation}
\begin{split}
 \mathcal K_{xc}(Q;a_x)={}&\mathcal K_x^{\mathrm{mSCAN}}(Q)
 +\mathcal K_c^{\mathrm{modSCAN}}(Q)\\
 &+a_x\bigl[\mathcal K_x^{\mathrm{HF,SR}}(Q)
 -\mathcal K_x^{\mathrm{mSCAN,SR}}(Q)\bigr].
\end{split}
\label{eq:finiteQ-components}
\end{equation}
Each term denotes the curvature of the corresponding energy
component evaluated using the same perturbation with $\omega=0.11\,\mathrm{bohr}^{-1}$.
Thus, $a_x=0$ gives the mSCAN-exchange plus modified-SCAN-
correlation reference, while $a_x=0.10$ replaces 10\% of its
short-range semilocal exchange by orbital-dependent Fock exchange.

The SR HF exchange energy is
evaluated using occupied non-interacting Bloch orbitals
obtained under the weak periodic potential
$v_s(z)=2V_Q\cos(Qz)$, with the short-range Coulomb
interaction $\mathrm{erfc}(\omega R)/R$.
The corresponding curvature
$\mathcal K_x^{\mathrm{HF,SR}}(Q)$ is determined from the
quadratic change in the exchange energy for small
positive and negative perturbations at fixed mean
electron density. Similarly,
$\mathcal K_x^{\mathrm{mSCAN,SR}}(Q)$ is obtained from
the short-range exchange energy defined by the
constructed mSCAN exchange hole, using the same
perturbed density and orbitals. The difference
$\mathcal K_x^{\mathrm{HF,SR}}(Q)-
\mathcal K_x^{\mathrm{mSCAN,SR}}(Q)$
therefore measures the change in the quadratic
exchange-energy response arising from the replacement
of semilocal SR exchange by Fock exchange.
These calculations do not include self-consistent
hybrid orbital relaxation.

The correlation contribution in Eq.~\eqref{eq:finiteQ-components}
comes from the modified SCAN correlation defined by Eq. (27) and Eq.(28)
%Eqs.~\eqref{eq:modSCAN-correlation} and
%\eqref{eq:modSCAN-interpolation}
. In the uniform gas $\alpha=1$.
The correlation interpolation factor $f_c^{\mathrm{int}}$ and
all its derivatives vanish at $\alpha=1$, so changing $c_{1c}$
from $0.64$ to $1.60$ does not alter the quadratic response
about this limit. Only the $\epsilon_c^1$ branch contributes.
For an unpolarized gas, its relevant expansion is
~\cite{scan2015,pw1992}
\begin{equation}
 \epsilon_c^1
 =\epsilon_c^{\mathrm{PW92}}(n)+\beta(r_s)t^2+O(t^4),
 \qquad t=\frac{|\nabla n|}{2k_s n},
 \label{eq:finiteQ-correlation-expansion}
\end{equation}
where $\epsilon_c^{\mathrm{PW92}}(n)$ is the uniform-gas
correlation energy per electron, $t$ is the reduced correlation
gradient, $k_s^2=4k_F/\pi$ is the Thomas--Fermi screening
wave vector squared, and
$\beta(r_s)=0.066725(1+0.1r_s)/(1+0.1778r_s)$~\cite{pw1992,pbe1996,scan2015,pbesol,tpss,revtpss}.

Substitution of Eq.~\eqref{eq:finiteQ-modulation} yields
\begin{equation}
 \mathcal K_c^{\mathrm{modSCAN}}(Q)
 =\left.\frac{d^2[n\epsilon_c^{\mathrm{PW92}}(n)]}{dn^2}
 \right|_{n=n_0}
 +\frac{\beta(r_s)Q^2}{2n_0k_s^2}.
 \label{eq:finiteQ-correlation-curvature}
\end{equation}
The first term is the uniform-gas correlation curvature;
the second is the quadratic contribution of the density
gradient. Equation~\eqref{eq:finiteQ-correlation-curvature}
is therefore the infinitesimal-modulation limit of the same
modified correlation functional employed in HSE-mSCAN,
not an independently selected correlation functional.
The perturbation amplitude tends to zero at fixed $Q$;
terms beyond quadratic order in the density modulation
are excluded.

We compare Eq.~\eqref{eq:finiteQ-Gcalc} with the analytical
local-field-factor parametrization of Corradini
\textit{et al.}~\cite{corradini1998}, denoted by $G_{\mathrm C}(Q)$.
This reference was constructed from diffusion Monte Carlo
results of Moroni \textit{et al.}~\cite{moroni1995} together
with known wave-vector limits. The values used here at
$r_s=2.07$, $3.0$, and $4.0$ are evaluations of the Corradini
parametrization~\cite{corradini1998}, not separately calculated QMC data at
these densities.

\begin{figure*}[t]
 \centering
 \includegraphics[width=0.96\textwidth]{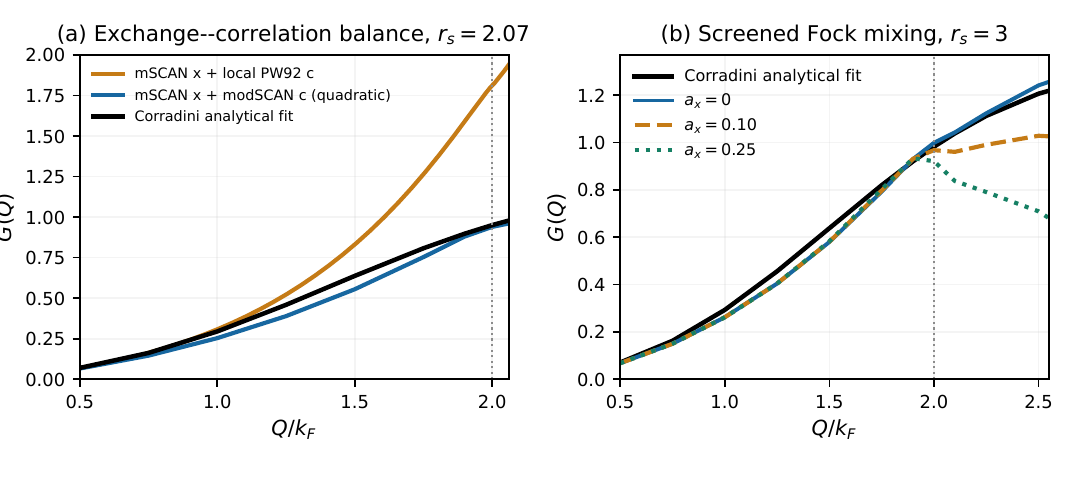}
 \caption{Finite-wave-vector XC energy-curvature test for the
 spin-unpolarized uniform electron gas. The horizontal axis is
 $Q/k_F$, where $Q$ is the density-modulation wave vector and
 $k_F$ is the Fermi wave vector of the unperturbed gas.
 (a) At $r_s=2.07$, the calculated $G_{\mathrm{calc}}$
 from mSCAN exchange plus local PW92 correlation is compared
 with the result including the quadratic gradient contribution
 of modified SCAN correlation.
 (b) At $r_s=3$, $G_{\mathrm{calc}}$ is shown for short-range
 Fock fractions $a_x=0$, $0.10$, and $0.25$ at fixed
 $\omega=0.11\,\mathrm{bohr}^{-1}$.
 Vertical dotted lines mark $Q=2k_F$.
 Black curves show the Corradini analytical parametrization
 $G_{\mathrm C}$~\cite{corradini1998}, based on earlier
 diffusion Monte Carlo data~\cite{moroni1995}, rather than
 raw QMC points at the plotted densities. The other curves
 are quadratic XC energy-curvature diagnostics obtained
 from perturbed noninteracting orbitals, not self-consistent
 generalized-Kohn--Sham response functions.}
 \label{fig:finiteQ-explained}
\end{figure*}

Figure~\ref{fig:finiteQ-explained} presents the two distinct
aspects of the test: panel (a) isolates the role of correlation
in the semilocal response, whereas panel (b) examines the
short-range Fock replacement.

Figure~\ref{fig:finiteQ-explained}(a) shows that correlation
has a large effect on the finite-wave-vector behavior.
At $r_s=2.07$ and $Q=2k_F$, mSCAN exchange combined
with local PW92 correlation gives $G_{\mathrm{calc}}=1.8110$.
Including the gradient contribution in
Eq.~\eqref{eq:finiteQ-correlation-curvature} reduces it to
$0.9395$, close to the Corradini value $G_{\mathrm C}=0.9503$.
This is a direct example of compensation between exchange
and correlation in the quadratic energy response; evaluating
the exchange contribution alone would not reveal it.

Figure~\ref{fig:finiteQ-explained}(b) shows how the
short-range Fock fraction changes this balance. The curves
are close for $Q\lesssim k_F$ and separate more strongly
as $Q$ approaches and exceeds $2k_F$. For $r_s=3$ at
$Q=2k_F$, the values are $0.9985$, $0.9675$, and $0.9209$
for $a_x=0$, $0.10$, and $0.25$, respectively, compared
with $G_{\mathrm C}=0.9820$. The 10\% fraction is marginally
closer to this reference than the unmixed parent and more
accurate than 25\% at this density and wave vector. At
larger $Q$, increasing Fock mixing does not systematically
improve the agreement.

\begin{table}
 \centering
 \caption{Calculated local-field-factor representation
 $G_{\mathrm{calc}}(Q;a_x)$ at $Q=2k_F$ and Corradini
 analytical reference $G_{\mathrm C}(Q)$ for three uniform-gas
 densities. The short-range screening parameter is
 $\omega=0.11\,\mathrm{bohr}^{-1}$. The reference values are
 evaluated from Ref.~\cite{corradini1998}, not direct QMC
 calculations at the listed densities.}
 \label{tab:finiteQ-response}
 \footnotesize
 \setlength{\tabcolsep}{3.0pt}
 \begin{tabular}{lrrrr}
 \toprule
 $r_s$ & $G_{\mathrm C}$ & $a_x=0$ & $a_x=0.10$ & $a_x=0.25$\\
 \midrule
 2.07 & 0.9503 & 0.9395 & 0.9117 & 0.8700\\
 3.00 & 0.9820 & 0.9985 & 0.9675 & 0.9209\\
 4.00 & 0.9999 & 1.0517 & 1.0190 & 0.9699\\
 \bottomrule
 \end{tabular}
\end{table}

Table~\ref{tab:finiteQ-response} shows the density dependence
of this comparison. At $Q=2k_F$, 10\% Fock mixing is closer
to the Corradini reference than 25\% for all three densities,
but the semilocal result is closest at $r_s=2.07$.
Consequently, the test motivates keeping the short-range
exchange replacement modest; it does not identify
$a_x=0.10$ as a unique or nonempirical optimum.

The semilocal contributions above use their quadratic
uniform-gas limits, whereas the short-range Fock curvature
is estimated from small but finite perturbations. Numerical
convergence, particularly near $Q=2k_F$, and the absence of
self-consistent hybrid orbital relaxation limit the strength
of the comparison with the interacting reference.

\section{Statistical definitions}\label{sec:si-statistics}

For a reference value $X_i^{\mathrm{ref}}$ and a calculated value $X_i$, the absolute error and relative error are defined as
\begin{equation}
 e_i=X_i-X_i^{\mathrm{ref}},\qquad
 r_i=100\frac{e_i}{X_i^{\mathrm{ref}}}.
\end{equation}
The reported statistical measures are
\begin{align}
 \mathrm{ME}&=\frac{1}{N}\sum_i e_i,
 &\mathrm{MAE}&=\frac{1}{N}\sum_i |e_i|,\\
 \mathrm{STDE}&=\left[\frac{1}{N}\sum_i(e_i-\mathrm{ME})^2\right]^{1/2},\\
 \mathrm{MRE}&=\frac{1}{N}\sum_i r_i,
 &\mathrm{MARE}&=\frac{1}{N}\sum_i |r_i|,\\
 \mathrm{STDRE}&=\left[\frac{1}{N}\sum_i(r_i-\mathrm{MRE})^2\right]^{1/2}.
\end{align}
The standard deviations follow the population convention, with normalization by $N$ rather than $N-1$. The numerical values reported in the tables reproduce the corresponding main-text statistics to the displayed precision. Relative errors can become large for systems with small reference values, even when the corresponding absolute errors are modest. Throughout, errors are defined as calculated minus reference values. Cohesive energies are expressed as positive binding energies per atom. Molecular atomization, barrier, and interaction energies are reported in kcal/mol, while molecular bond-length errors are given in m\AA, as specified in the main text.

\section{Complete solid-state benchmark datasets}

This section provides the complete solid-state benchmark datasets used in the main text. The structural and cohesive-property datasets comprise 46 cubic solids, while the band-gap dataset contains 76 structure-resolved systems. The calculated values are reproduced from the numerical compilation, together with the corresponding experimental references and crystallographic space groups. No spin--orbit coupling (SOC) or empirical SOC correction to the band gaps is included. The adopted experimental references originate from different compilations and may correspond to different measurement temperatures and correction conventions. Consequently, they do not constitute a uniform set of static-lattice reference values~\cite{haas2009,haas2009erratum,schimka2011,csonka2009,Kaplan2022,Janthon2014,Zhang2018,peverati2012,tao2017}.

The Y and N flags reproduce the correction classifications in the numerical compilation. For lattice constants, Y indicates that a zero-point anharmonic expansion (ZPAE) treatment is adopted. For bulk moduli and cohesive energies, Y indicates that a zero-point correction is adopted. A dagger marks a correction declaration inherited from the original compilation, whereas a double dagger identifies a retained reference for which the correction was not independently verified. In some cases, changing the reference source also changes the experimental baseline; such a change should not be interpreted as a pure zero-point correction. For band gaps, N indicates that no explicit zero-point renormalization (ZPR) adjustment is applied. It does not imply that the experimental reference is free from thermal or electron--phonon effects. These flags preserve the classifications in the numerical compilation and do not represent an independent re-evaluation of the corrections for each material.

For the class-resolved structural statistics, the simple/$sp$-metal group comprises Al, Ba, Ca, Cs, K, Li, Na, Pb, Rb, and Sr. The transition-metal group comprises Ag, Au, Cu, Fe, Ir, Mo, Nb, Ni, Pd, Pt, Rh, Ta, V, and W. The insulator group comprises AlN, BN, LiCl, LiF, MgO, NaCl, and NaF. The remaining 15 materials are classified as semiconductors. These group assignments define the classification used to calculate and plot the class-resolved statistics; they do not imply a classification of the magnetic states.

\subsection{Lattice constants}\footnotesize\setlength{\tabcolsep}{3pt}
\begin{longtable}{llrrrrll}\caption{Complete lattice-constant dataset (\AA). The correction column refers to the adopted zero-point anharmonic-expansion (ZPAE) treatment.}\label{tab:full-lattice}\\\toprule Material & Space group & Ref. value & SCAN & HSE06 & HSE-mSCAN & \makecell{ZPE\\corrected?} & Reference\\\midrule\endfirsthead
\toprule Material & Space group & Ref. value & SCAN & HSE06 & HSE-mSCAN & \makecell{ZPE\\corrected?} & Reference\\\midrule\endhead
Ag & $Fm\overline{3}m$ (225) & 4.0620 & 4.0789 & 4.1429 & 4.1182 & Y & \cite{haas2009}\\
Al & $Fm\overline{3}m$ (225) & 4.0190 & 4.0028 & 4.0290 & 3.9733 & Y & \cite{haas2009}\\
AlAs & $F\overline{4}3m$ (216) & 5.6460 & 5.6723 & 5.6765 & 5.6634 & Y & \cite{haas2009}\\
AlN & $F\overline{4}3m$ (216) & 4.3680 & 4.3630 & 4.3644 & 4.3420 & Y & \cite{schimka2011}\\
AlP & $F\overline{4}3m$ (216) & 5.4450 & 5.4660 & 5.4707 & 5.4611 & Y & \cite{haas2009}\\
AlSb & $F\overline{4}3m$ (216) & 6.1260 & 6.1726 & 6.1522 & 6.1656 & Y & \cite{peverati2012}\\
Au & $Fm\overline{3}m$ (225) & 4.0620 & 4.0887 & 4.1288 & 4.1217 & Y & \cite{haas2009}\\
Ba & $Im\overline{3}m$ (229) & 5.0020 & 5.0405 & 5.1225 & 5.1582 & Y & \cite{haas2009}\\
BAs & $F\overline{4}3m$ (216) & 4.7600 & 4.7778 & 4.7705 & 4.7695 & Y & \cite{haas2009}\\
BN & $F\overline{4}3m$ (216) & 3.5850 & 3.6063 & 3.5979 & 3.5905 & Y & \cite{haas2009}\\
BP & $F\overline{4}3m$ (216) & 4.5200 & 4.5200 & 4.5200 & 4.5200 & Y & \cite{haas2009}\\
C & $Fd\overline{3}m$ (227) & 3.5440 & 3.5543 & 3.5471 & 3.5506 & Y & \cite{haas2009}\\
Ca & $Fm\overline{3}m$ (225) & 5.5560 & 5.5504 & 5.5960 & 5.5560 & Y & \cite{haas2009}\\
Cs & $Im\overline{3}m$ (229) & 6.0390 & 6.2211 & 6.2387 & 6.3173 & Y & \cite{tao2017}\\
Cu & $Fm\overline{3}m$ (225) & 3.5960 & 3.5860 & 3.6358 & 3.5960 & Y & \cite{haas2009}\\
Fe & $Im\overline{3}m$ (229) & 2.8530 & 2.8435 & 2.9060 & 2.8806 & Y & \cite{haas2009}\\
GaAs & $F\overline{4}3m$ (216) & 5.6370 & 5.6647 & 5.6672 & 5.6549 & Y & \cite{haas2009}\\
GaN & $F\overline{4}3m$ (216) & 4.5200 & 4.5267 & 4.5215 & 4.5022 & Y & \cite{haas2009}\\
GaP & $F\overline{4}3m$ (216) & 5.4350 & 5.4627 & 5.4636 & 5.4526 & Y & \cite{haas2009}\\
Ge & $Fd\overline{3}m$ (227) & 5.6390 & 5.6841 & 5.6755 & 5.6671 & Y & \cite{haas2009}\\
InAs & $F\overline{4}3m$ (216) & 6.0440 & 6.1216 & 6.1090 & 6.1127 & Y & \cite{haas2009}\\
InP & $F\overline{4}3m$ (216) & 5.8560 & 5.9355 & 5.9223 & 5.9207 & Y & \cite{haas2009}\\
InSb & $F\overline{4}3m$ (216) & 6.4730 & 6.5449 & 6.5196 & 6.5474 & Y & \cite{schimka2011}\\
Ir & $Fm\overline{3}m$ (225) & 3.8310 & 3.7878 & 3.8302 & 3.7956 & Y & \cite{haas2009}\\
K & $Im\overline{3}m$ (229) & 5.2120 & 5.3046 & 5.3107 & 5.3649 & Y & \cite{haas2009}\\
Li & $Im\overline{3}m$ (229) & 3.4510 & 3.4638 & 3.4679 & 3.4644 & Y & \cite{haas2009}\\
LiCl & $Fm\overline{3}m$ (225) & 5.0720 & 5.0910 & 5.1167 & 5.1081 & Y & \cite{haas2009}\\
LiF & $Fm\overline{3}m$ (225) & 3.9600 & 3.9772 & 4.0197 & 3.9501 & Y & \cite{haas2009}\\
MgO & $Fm\overline{3}m$ (225) & 4.1860 & 4.1860 & 4.1984 & 4.1748 & Y & \cite{haas2009}\\
Mo & $Im\overline{3}m$ (229) & 3.1410 & 3.1367 & 3.1293 & 3.1380 & Y & \cite{haas2009}\\
Na & $Im\overline{3}m$ (229) & 4.2090 & 4.1853 & 4.2090 & 4.2090 & Y & \cite{haas2009}\\
NaCl & $Fm\overline{3}m$ (225) & 5.5650 & 5.5211 & 5.6140 & 5.5468 & Y & \cite{haas2009}\\
NaF & $Fm\overline{3}m$ (225) & 4.5760 & 4.4735 & 4.5760 & 4.4807 & Y & \cite{haas2009}\\
Nb & $Im\overline{3}m$ (229) & 3.2940 & 3.2921 & 3.2836 & 3.2966 & Y & \cite{haas2009}\\
Ni & $Fm\overline{3}m$ (225) & 3.5080 & 3.4543 & 3.5080 & 3.4645 & Y & \cite{haas2009}\\
Pb & $Fm\overline{3}m$ (225) & 4.9120 & 4.9710 & 4.9945 & 4.9937 & Y & \cite{haas2009}\\
Pd & $Fm\overline{3}m$ (225) & 3.8760 & 3.8920 & 3.9071 & 3.9134 & Y & \cite{haas2009}\\
Pt & $Fm\overline{3}m$ (225) & 3.9130 & 3.8969 & 3.9284 & 3.9137 & Y & \cite{haas2009}\\
Rb & $Im\overline{3}m$ (229) & 5.5770 & 5.6975 & 5.7154 & 5.7870 & Y & \cite{haas2009}\\
Rh & $Fm\overline{3}m$ (225) & 3.7930 & 3.7778 & 3.7755 & 3.7865 & Y & \cite{haas2009}\\
Si & $Fd\overline{3}m$ (227) & 5.4150 & 5.4269 & 5.4332 & 5.4226 & Y & \cite{haas2009}\\
SiC & $F\overline{4}3m$ (216) & 4.3400 & 4.3517 & 4.3463 & 4.3414 & Y & \cite{haas2009}\\
Sr & $Fm\overline{3}m$ (225) & 6.0400 & 6.0837 & 6.1165 & 6.1231 & Y & \cite{haas2009}\\
Ta & $Im\overline{3}m$ (229) & 3.2990 & 3.2699 & 3.2854 & 3.2660 & Y & \cite{haas2009}\\
V & $Im\overline{3}m$ (229) & 3.0240 & 2.9518 & 2.9470 & 2.9540 & Y & \cite{haas2009}\\
W & $Im\overline{3}m$ (229) & 3.1600 & 3.1404 & 3.1427 & 3.1401 & Y & \cite{haas2009}\\
\bottomrule\end{longtable}

\subsection{Bulk moduli}\footnotesize\setlength{\tabcolsep}{3pt}
\begin{longtable}{llrrrrll}\caption{Complete bulk-modulus dataset (GPa).}\label{tab:full-bulk}\\\toprule Material & Space group & Ref. value & SCAN & HSE06 & HSE-mSCAN & \makecell{ZPE\\corrected?} & Reference\\\midrule\endfirsthead
\toprule Material & Space group & Ref. value & SCAN & HSE06 & HSE-mSCAN & \makecell{ZPE\\corrected?} & Reference\\\midrule\endhead
Ag & $Fm\overline{3}m$ (225) & 110.800 & 109.666 & 87.275 & 94.336 & Y & \cite{schimka2011}\\
Al & $Fm\overline{3}m$ (225) & 82.000 & 77.742 & 74.525 & 83.416 & Y & \cite{schimka2011}\\
AlAs & $F\overline{4}3m$ (216) & 75.000 & 73.798 & 73.769 & 76.735 & Y & \cite{schimka2011}\\
AlN & $F\overline{4}3m$ (216) & 206.000 & 213.204 & 214.754 & 219.919 & Y & \cite{schimka2011}\\
AlP & $F\overline{4}3m$ (216) & 87.400 & 88.243 & 90.896 & 90.531 & Y & \cite{schimka2011}\\
AlSb & $F\overline{4}3m$ (216) & 59.300 & 54.256 & 55.494 & 56.532 & N & \cite{Zhang2018}\\
Au & $Fm\overline{3}m$ (225) & 174.800 & 158.291 & 144.380 & 151.352 & Y & \cite{Janthon2014}\\
Ba & $Im\overline{3}m$ (229) & 10.600 & 8.305 & 8.273 & 8.028 & Y & \cite{Kaplan2022}\\
BAs & $F\overline{4}3m$ (216) & 151.100 & 145.542 & 146.674 & 143.670 & Y & \cite{schimka2011}\\
BN & $F\overline{4}3m$ (216) & 410.200 & 386.392 & 393.100 & 390.361 & Y & \cite{schimka2011}\\
BP & $F\overline{4}3m$ (216) & 168.000 & 173.928 & 169.242 & 169.427 & Y & \cite{schimka2011}\\
C & $Fd\overline{3}m$ (227) & 454.700 & 453.291 & 464.257 & 464.635 & Y & \cite{schimka2011}\\
Ca & $Fm\overline{3}m$ (225) & 15.900 & 17.137 & 16.786 & 18.959 & Y & \cite{Kaplan2022}\\
Cs & $Im\overline{3}m$ (229) & 2.300 & 1.919 & 1.963 & 1.986 & Y & \cite{Kaplan2022}\\
Cu & $Fm\overline{3}m$ (225) & 145.000 & 154.942 & 127.042 & 133.651 & Y & \cite{schimka2011}\\
Fe & $Im\overline{3}m$ (229) & 169.800 & 165.047 & 175.831 & 176.295 & Y & \cite{Janthon2014}\\
GaAs & $F\overline{4}3m$ (216) & 76.700 & 71.480 & 72.479 & 75.679 & Y & \cite{schimka2011}\\
GaN & $F\overline{4}3m$ (216) & 213.700 & 212.446 & 215.464 & 226.477 & Y & \cite{schimka2011}\\
GaP & $F\overline{4}3m$ (216) & 89.600 & 87.699 & 88.425 & 91.454 & Y & \cite{schimka2011}\\
Ge & $Fd\overline{3}m$ (227) & 77.300 & 68.944 & 71.051 & 74.066 & Y & \cite{schimka2011}\\
InAs & $F\overline{4}3m$ (216) & 58.600 & 56.804 & 58.072 & 59.802 & Y & \cite{schimka2011}\\
InP & $F\overline{4}3m$ (216) & 72.000 & 68.179 & 72.161 & 71.587 & Y & \cite{schimka2011}\\
InSb & $F\overline{4}3m$ (216) & 46.100 & 42.099 & 44.401 & 44.306 & Y & \cite{schimka2011}\\
Ir & $Fm\overline{3}m$ (225) & 365.200 & 409.937 & 395.571 & 388.634 & Y & \cite{Janthon2014}\\
K & $Im\overline{3}m$ (229) & 3.800 & 3.429 & 3.421 & 3.382 & Y & \cite{Kaplan2022}\\
Li & $Im\overline{3}m$ (229) & 13.900 & 12.966 & 13.167 & 13.437 & Y & \cite{schimka2011}\\
LiCl & $Fm\overline{3}m$ (225) & 38.700 & 35.043 & 32.699 & 34.323 & Y & \cite{schimka2011}\\
LiF & $Fm\overline{3}m$ (225) & 76.300 & 78.847 & 72.168 & 76.598 & Y & \cite{schimka2011}\\
MgO & $Fm\overline{3}m$ (225) & 169.800 & 167.945 & 165.441 & 172.410 & Y & \cite{schimka2011}\\
Mo & $Im\overline{3}m$ (229) & 265.300 & 285.534 & 283.706 & 280.261 & Y & \cite{Zhang2018}\\
Na & $Im\overline{3}m$ (229) & 7.700 & 7.826 & 7.545 & 7.534 & Y & \cite{schimka2011}\\
NaCl & $Fm\overline{3}m$ (225) & 27.600 & 29.124 & 25.068 & 27.628 & Y & \cite{schimka2011}\\
NaF & $Fm\overline{3}m$ (225) & 53.100 & 60.537 & 50.675 & 58.307 & Y & \cite{schimka2011}\\
Nb & $Im\overline{3}m$ (229) & 172.000 & 172.127 & 170.159 & 166.083 & Y & \cite{Janthon2014}\\
Ni & $Fm\overline{3}m$ (225) & 185.500 & 209.707 & 121.738 & 198.470 & Y & \cite{Janthon2014}\\
Pb & $Fm\overline{3}m$ (225) & 47.300 & 44.746 & 44.516 & 44.420 & Y & \cite{csonka2009}\\
Pd & $Fm\overline{3}m$ (225) & 198.100 & 186.265 & 166.283 & 173.527 & Y & \cite{schimka2011}\\
Pt & $Fm\overline{3}m$ (225) & 284.200 & 293.735 & 269.954 & 280.248 & Y & \cite{Janthon2014}\\
Rb & $Im\overline{3}m$ (229) & 3.600 & 2.643 & 2.709 & 3.242 & Y & \cite{Kaplan2022}\\
Rh & $Fm\overline{3}m$ (225) & 272.100 & 278.253 & 295.906 & 285.587 & Y & \cite{schimka2011}\\
Si & $Fd\overline{3}m$ (227) & 100.800 & 99.257 & 94.898 & 96.698 & Y & \cite{schimka2011}\\
SiC & $F\overline{4}3m$ (216) & 229.100 & 226.758 & 232.860 & 233.322 & Y & \cite{schimka2011}\\
Sr & $Fm\overline{3}m$ (225) & 12.000 & 11.253 & 10.926 & 12.455 & Y & \cite{Kaplan2022}\\
Ta & $Im\overline{3}m$ (229) & 193.700 & 208.318 & 231.240 & 208.343 & Y & \cite{Janthon2014}\\
V & $Im\overline{3}m$ (229) & 158.900 & 198.487 & 220.357 & 211.202 & Y & \cite{Janthon2014}\\
W & $Im\overline{3}m$ (229) & 312.300 & 340.286 & 353.056 & 350.900 & Y & \cite{Janthon2014}\\
\bottomrule\end{longtable}

\subsection{Complete cohesive energies (eV/atom)}\footnotesize\setlength{\tabcolsep}{3pt}
\begin{longtable}{llrrrrll}\caption{Complete cohesive energies dataset (eV/atom). }\label{tab:full-cohesive}\\\toprule Material & Space group & Ref. value & SCAN & HSE06 & HSE-mSCAN & \makecell{ZPE\\corrected?} & Reference\\\midrule\endfirsthead
\toprule Material & Space group & Ref. value & SCAN & HSE06 & HSE-mSCAN & \makecell{ZPE\\corrected?} & Reference\\\midrule\endhead
Ag & $Fm\overline{3}m$ (225) & 2.9600 & 2.8800 & 2.3650 & 2.5966 & Y & \cite{Kaplan2022}\\
Al & $Fm\overline{3}m$ (225) & 3.4300 & 3.6030 & 3.4180 & 3.4492 & Y & \cite{Kaplan2022}\\
AlAs & $F\overline{4}3m$ (216) & 3.8244 & 3.8950 & 3.7090 & 3.8047 & Y & \cite{schimka2011}\\
AlN & $F\overline{4}3m$ (216) & 5.8454 & 5.8200 & 5.6270 & 5.7732 & Y & \cite{schimka2011}\\
AlP & $F\overline{4}3m$ (216) & 4.3219 & 4.2650 & 4.1110 & 4.2233 & Y & \cite{schimka2011}\\
AlSb & $F\overline{4}3m$ (216) & 3.3000 & 3.4220 & 3.2900 & 3.3221 & N & \cite{Zhang2018}\\
Au & $Fm\overline{3}m$ (225) & 3.8300 & 3.5540 & 2.8930 & 3.1525 & Y & \cite{Janthon2014}\\
Ba & $Im\overline{3}m$ (229) & 1.9100 & 2.0240 & 1.7650 & 1.8903 & Y & \cite{Kaplan2022}\\
BAs & $F\overline{4}3m$ (216) & 4.7000 & 4.7040 & 4.5940 & 4.5833 & N & \cite{Zhang2018}\\
BN & $F\overline{4}3m$ (216) & 6.7575 & 6.8340 & 6.8360 & 6.7531 & Y & \cite{schimka2011}\\
BP & $F\overline{4}3m$ (216) & 5.1407 & 5.3280 & 5.2720 & 5.2293 & Y & \cite{schimka2011}\\
C & $Fd\overline{3}m$ (227) & 7.5500 & 7.5010 & 7.6140 & 7.3827 & Y & \cite{Kaplan2022}\\
Ca & $Fm\overline{3}m$ (225) & 1.8700 & 2.0750 & 1.7940 & 1.9863 & Y & \cite{Kaplan2022}\\
Cs & $Im\overline{3}m$ (229) & 0.8100 & 0.6930 & 0.6920 & 0.6515 & Y & \cite{Kaplan2022}\\
Cu & $Fm\overline{3}m$ (225) & 3.5100 & 3.8800 & 3.0690 & 3.4688 & Y & \cite{Kaplan2022}\\
Fe & $Im\overline{3}m$ (229) & 4.3200 & 4.5880 & 3.3720 & 3.8618 & Y & \cite{Janthon2014}\\
GaAs & $F\overline{4}3m$ (216) & 3.3400 & 3.3690 & 3.1970 & 3.2232 & Y & \cite{Kaplan2022}\\
GaN & $F\overline{4}3m$ (216) & 4.5499 & 4.4640 & 4.3260 & 4.3868 & Y & \cite{schimka2011}\\
GaP & $F\overline{4}3m$ (216) & 3.6068 & 3.6570 & 3.5360 & 3.5842 & Y & \cite{schimka2011}\\
Ge & $Fd\overline{3}m$ (227) & 3.8900 & 3.9770 & 3.7540 & 3.8041 & Y & \cite{Kaplan2022}\\
InAs & $F\overline{4}3m$ (216) & 3.0782 & 3.0300 & 2.9040 & 2.8969 & Y & \cite{schimka2011}\\
InP & $F\overline{4}3m$ (216) & 3.4720 & 3.2320 & 3.1510 & 3.1795 & Y & \cite{schimka2011}\\
InSb & $F\overline{4}3m$ (216) & 2.8087 & 2.8050 & 2.6940 & 2.6406 & Y & \cite{schimka2011}\\
Ir & $Fm\overline{3}m$ (225) & 6.9600 & 7.6060 & 6.4480 & 6.6740 & Y & \cite{Janthon2014}\\
K & $Im\overline{3}m$ (229) & 0.9400 & 0.8620 & 0.8420 & 0.8278 & Y & \cite{Kaplan2022}\\
Li & $Im\overline{3}m$ (229) & 1.6700 & 1.5640 & 1.5480 & 1.5404 & Y & \cite{Kaplan2022}\\
LiCl & $Fm\overline{3}m$ (225) & 3.5900 & 3.4910 & 3.3570 & 3.4908 & Y & \cite{Kaplan2022}\\
LiF & $Fm\overline{3}m$ (225) & 4.4600 & 4.3960 & 4.2160 & 4.3712 & Y & \cite{Kaplan2022}\\
MgO & $Fm\overline{3}m$ (225) & 5.2000 & 5.2530 & 4.9070 & 5.2159 & Y & \cite{Kaplan2022}\\
Mo & $Im\overline{3}m$ (229) & 6.8400 & 5.9240 & 5.2270 & 5.1403 & Y & \cite{Janthon2014}\\
Na & $Im\overline{3}m$ (229) & 1.1200 & 1.1010 & 1.0470 & 1.0411 & Y & \cite{Kaplan2022}\\
NaCl & $Fm\overline{3}m$ (225) & 3.3400 & 3.3020 & 3.1170 & 3.2977 & Y & \cite{Kaplan2022}\\
NaF & $Fm\overline{3}m$ (225) & 3.9700 & 4.0140 & 3.7650 & 3.9760 & Y & \cite{Kaplan2022}\\
Nb & $Im\overline{3}m$ (229) & 7.4700 & 6.6610 & 6.1920 & 6.0514 & Y & \cite{Janthon2014}\\
Ni & $Fm\overline{3}m$ (225) & 4.4800 & 4.4300 & 3.2880 & 3.6351 & Y & \cite{Janthon2014}\\
Pb & $Fm\overline{3}m$ (225) & 2.0300 & 3.0750 & 2.7920 & 2.7932 & N$^{\ddagger}$ & \cite{Zhang2018}\\
Pd & $Fm\overline{3}m$ (225) & 3.9300 & 4.3750 & 2.9980 & 3.6879 & Y & \cite{Kaplan2022}\\
Pt & $Fm\overline{3}m$ (225) & 5.8700 & 5.9430 & 4.8780 & 5.1695 & Y & \cite{Janthon2014}\\
Rb & $Im\overline{3}m$ (229) & 0.8600 & 0.7590 & 0.7530 & 0.7286 & Y & \cite{Kaplan2022}\\
Rh & $Fm\overline{3}m$ (225) & 5.7800 & 5.4540 & 4.7030 & 5.2105 & Y & \cite{Kaplan2022}\\
Si & $Fd\overline{3}m$ (227) & 4.6800 & 4.7100 & 4.5850 & 4.6342 & Y & \cite{Kaplan2022}\\
SiC & $F\overline{4}3m$ (216) & 6.4800 & 6.4440 & 6.4100 & 6.4234 & Y & \cite{Kaplan2022}\\
Sr & $Fm\overline{3}m$ (225) & 1.7300 & 1.8060 & 1.5650 & 1.7164 & Y & \cite{Kaplan2022}\\
Ta & $Im\overline{3}m$ (229) & 8.1100 & 8.6210 & 7.8050 & 8.0644 & Y & \cite{Janthon2014}\\
V & $Im\overline{3}m$ (229) & 5.3400 & 4.7290 & 3.4570 & 3.7858 & Y & \cite{Janthon2014}\\
W & $Im\overline{3}m$ (229) & 8.8300 & 9.0060 & 8.4840 & 8.9759 & Y & \cite{Janthon2014}\\
\bottomrule\end{longtable}

\subsection{Complete band gaps}\footnotesize\setlength{\tabcolsep}{3pt}
\begin{longtable}{llrrrrll}\caption{Complete band-gap dataset (eV). No explicit zero-point band-gap renormalization (ZPR) is applied to the compiled experimental references. Space groups describe reference crystal structures, not magnetic symmetry.}\label{tab:full-gap}\\\toprule Material & Space group & Ref. value & SCAN & HSE06 & HSE-mSCAN & \makecell{ZPR\\corrected?} & Reference\\\midrule\endfirsthead
\toprule Material & Space group & Ref. value & SCAN & HSE06 & HSE-mSCAN & \makecell{ZPR\\corrected?} & Reference\\\midrule\endhead
AgBr & $Fm\overline{3}m$ (225) & 2.710 & 1.150 & 2.010 & 1.920 & N & \cite{tran2017}\\
AgCl & $Fm\overline{3}m$ (225) & 3.250 & 1.330 & 2.390 & 2.100 & N & \cite{tran2017}\\
AgI & $F\overline{4}3m$ (216) & 2.910 & 1.650 & 2.470 & 2.400 & N & \cite{tran2017}\\
Al$_{2}$O$_{3}$ & $R\overline{3}c$ (167) & 8.800 & 7.120 & 8.100 & 8.010 & N & \cite{tran2017}\\
AlAs & $F\overline{4}3m$ (216) & 2.230 & 1.740 & 2.100 & 2.190 & N & \cite{tran2017}\\
AlN & $F\overline{4}3m$ (216) & 4.900 & 3.930 & 4.580 & 4.570 & N & \cite{tran2017}\\
AlN (wurtzite) & $P6_{3}mc$ (186) & 6.190 & 4.800 & 5.510 & 5.530 & N & \cite{tran2017}\\
AlP & $F\overline{4}3m$ (216) & 2.500 & 1.920 & 2.280 & 2.320 & N & \cite{tran2017}\\
AlSb & $F\overline{4}3m$ (216) & 1.690 & 1.370 & 1.810 & 1.760 & N & \cite{tran2017}\\
Ar & $Fm\overline{3}m$ (225) & 14.150 & 9.540 & 10.330 & 10.410 & N & \cite{tran2017}\\
BaS & $Fm\overline{3}m$ (225) & 3.880 & 2.460 & 3.020 & 2.940 & N & \cite{tran2017}\\
BAs & $F\overline{4}3m$ (216) & 1.460 & 1.420 & 1.850 & 1.770 & N & \cite{tran2017}\\
BaSe & $Fm\overline{3}m$ (225) & 3.580 & 2.250 & 2.720 & 2.720 & N & \cite{tran2017}\\
BaTe & $Fm\overline{3}m$ (225) & 3.080 & 1.860 & 2.250 & 2.310 & N & \cite{tran2017}\\
BeO (wurtzite) & $P6_{3}mc$ (186) & 10.600 & 8.500 & 9.500 & 9.400 & N & \cite{tran2017}\\
BN & $F\overline{4}3m$ (216) & 6.360 & 4.930 & 5.780 & 5.570 & N & \cite{tran2017}\\
BP & $F\overline{4}3m$ (216) & 2.100 & 1.540 & 1.970 & 1.830 & N & \cite{tran2017}\\
C & $Fd\overline{3}m$ (227) & 5.500 & 4.570 & 5.320 & 4.960 & N & \cite{tran2017}\\
CaF$_{2}$ & $Fm\overline{3}m$ (225) & 11.800 & 7.890 & 9.340 & 8.910 & N & \cite{tran2017}\\
CaO & $Fm\overline{3}m$ (225) & 7.000 & 4.160 & 5.300 & 4.900 & N & \cite{tran2017}\\
CdS & $F\overline{4}3m$ (216) & 2.500 & 1.530 & 2.190 & 2.160 & N & \cite{tran2017}\\
CdSe & $F\overline{4}3m$ (216) & 1.850 & 1.010 & 1.590 & 1.680 & N & \cite{tran2017}\\
CdTe & $F\overline{4}3m$ (216) & 1.610 & 1.040 & 1.550 & 1.760 & N & \cite{tran2017}\\
Cu$_{2}$O & $Pn\overline{3}m$ (224) & 2.170 & 0.780 & 2.000 & 1.370 & N & \cite{tran2017}\\
CuBr & $F\overline{4}3m$ (216) & 3.070 & 0.670 & 2.070 & 1.530 & N & \cite{tran2017}\\
CuCl & $F\overline{4}3m$ (216) & 3.400 & 0.720 & 2.280 & 1.560 & N & \cite{tran2017}\\
CuI & $F\overline{4}3m$ (216) & 3.120 & 1.470 & 2.580 & 2.350 & N & \cite{tran2017}\\
CuSCN & $R3m$ (160) & 3.940 & 2.450 & 3.560 & 3.100 & N & \cite{tran2017}\\
GaAs & $F\overline{4}3m$ (216) & 1.520 & 0.860 & 1.450 & 1.570 & N & \cite{tran2017}\\
GaN & $F\overline{4}3m$ (216) & 3.280 & 2.110 & 2.960 & 2.760 & N & \cite{tran2017}\\
GaN (wurtzite) & $P6_{3}mc$ (186) & 3.500 & 2.380 & 3.250 & 3.060 & N & \cite{tran2017}\\
GaP & $F\overline{4}3m$ (216) & 2.350 & 1.860 & 2.290 & 2.280 & N & \cite{tran2017}\\
GaSb & $F\overline{4}3m$ (216) & 0.820 & 0.200 & 0.930 & 0.920 & N & \cite{tran2017}\\
Ge & $Fd\overline{3}m$ (227) & 0.740 & 0.230 & 0.830 & 0.850 & N & \cite{tran2017}\\
InAs & $F\overline{4}3m$ (216) & 0.420 & 0.060 & 0.540 & 0.690 & N & \cite{tran2017}\\
InN (wurtzite) & $P6_{3}mc$ (186) & 0.720 & 0.130 & 0.800 & 0.630 & N & \cite{tran2017}\\
InP & $F\overline{4}3m$ (216) & 1.420 & 1.090 & 1.520 & 1.630 & N & \cite{tran2017}\\
InSb & $F\overline{4}3m$ (216) & 0.240 & 0.000 & 0.540 & 0.550 & N & \cite{tran2017}\\
KCl & $Fm\overline{3}m$ (225) & 8.500 & 5.770 & 6.500 & 6.480 & N & \cite{tran2017}\\
KF & $Fm\overline{3}m$ (225) & 10.900 & 6.750 & 8.160 & 7.720 & N & \cite{tran2017}\\
Kr & $Fm\overline{3}m$ (225) & 11.590 & 8.030 & 8.650 & 8.850 & N & \cite{tran2017}\\
LiCl & $Fm\overline{3}m$ (225) & 9.400 & 7.300 & 7.800 & 7.990 & N & \cite{tran2017}\\
LiF & $Fm\overline{3}m$ (225) & 14.200 & 10.120 & 11.500 & 11.190 & N & \cite{tran2017}\\
LiH & $Fm\overline{3}m$ (225) & 4.940 & 3.580 & 3.990 & 4.250 & N & \cite{tran2017}\\
MgO & $Fm\overline{3}m$ (225) & 7.830 & 5.600 & 6.480 & 6.440 & N & \cite{tran2017}\\
MgS & $F\overline{4}3m$ (216) & 4.780 & 4.230 & 4.630 & 4.810 & N & \cite{tran2017}\\
MgSe & $Fm\overline{3}m$ (225) & 2.470 & 2.480 & 2.710 & 3.000 & N & \cite{tran2017}\\
MgTe & $F\overline{4}3m$ (216) & 3.600 & 3.110 & 3.260 & 3.650 & N & \cite{tran2017}\\
MoS$_{2}$ & $P6_{3}/mmc$ (194) & 1.290 & 0.990 & 1.440 & 1.200 & N & \cite{tran2017}\\
NaCl & $Fm\overline{3}m$ (225) & 8.500 & 5.930 & 6.530 & 6.640 & N & \cite{tran2017}\\
NaF & $Fm\overline{3}m$ (225) & 11.500 & 7.100 & 8.490 & 8.130 & N & \cite{tran2017}\\
Ne & $Fm\overline{3}m$ (225) & 21.480 & 12.830 & 14.320 & 14.130 & N & \cite{tran2017}\\
Sb$_{2}$Te$_{3}$ & $R\overline{3}m$ (166) & 0.280 & 0.010 & 0.210 & 0.260 & N & \cite{tran2017}\\
ScN & $Fm\overline{3}m$ (225) & 0.900 & 0.270 & 0.820 & 0.700 & N & \cite{tran2017}\\
Si & $Fd\overline{3}m$ (227) & 1.170 & 0.820 & 1.150 & 1.070 & N & \cite{tran2017}\\
SiC & $F\overline{4}3m$ (216) & 2.420 & 1.730 & 2.240 & 2.230 & N & \cite{tran2017}\\
SiO$_{2}$ (alpha-quartz) & $P3_{1}21$ (152) & 9.650 & 6.490 & 7.770 & 7.420 & N & \cite{tran2017}\\
SiO$_{2}$ (beta-cristobalite) & $Fd\overline{3}m$ (227) & 8.900 & 6.090 & 7.330 & 6.960 & N & \cite{tran2017}\\
SnO$_{2}$ & $P4_{2}/mnm$ (136) & 3.590 & 1.800 & 2.940 & 2.610 & N & \cite{tran2017}\\
SnSe & $Pnma$ (62) & 0.900 & 0.740 & 0.980 & 1.060 & N & \cite{tran2017}\\
SnTe & $Fm\overline{3}m$ (225) & 0.360 & 0.100 & 0.140 & 0.200 & N & \cite{tran2017}\\
SrTiO$_{3}$ & $Pm\overline{3}m$ (221) & 3.300 & 2.100 & 3.090 & 2.640 & N & \cite{tran2017}\\
TiO$_{2}$ (anatase) & $I4_{1}/amd$ (141) & 3.400 & 2.300 & 3.310 & 2.840 & N & \cite{tran2017}\\
TiO$_{2}$ (rutile) & $P4_{2}/mnm$ (136) & 3.300 & 2.040 & 3.110 & 2.570 & N & \cite{tran2017}\\
VO$_{2}$ & $P2_{1}/c$ (14) & 0.600 & 0.000 & 0.940 & 0.280 & N & \cite{tran2017}\\
Xe & $Fm\overline{3}m$ (225) & 9.290 & 6.920 & 7.370 & 7.650 & N & \cite{tran2017}\\
ZnO (wurtzite) & $P6_{3}mc$ (186) & 3.440 & 1.150 & 2.460 & 1.950 & N & \cite{tran2017}\\
ZnS & $F\overline{4}3m$ (216) & 3.840 & 2.610 & 3.310 & 3.290 & N & \cite{tran2017}\\
ZnSe & $F\overline{4}3m$ (216) & 2.820 & 1.760 & 2.410 & 2.530 & N & \cite{tran2017}\\
ZnTe & $F\overline{4}3m$ (216) & 2.390 & 1.600 & 2.170 & 2.410 & N & \cite{tran2017}\\
CoO & $Fm\overline{3}m$ (225) & 2.500 & 1.010 & 3.590 & 2.430 & N & \cite{tran2017}\\
FeO & $Fm\overline{3}m$(225)& 2.400 & 0.540 & 2.360 & 1.330 & N & \cite{tran2017}\\
MnO & $Fm\overline{3}m$ (225)& 3.900 & 1.630 & 2.950 & 2.590 & N & \cite{tran2017}\\
NiO & $Fm\overline{3}m$ (225)& 4.300 & 2.450 & 4.560 & 3.820 & N & \cite{tran2017}\\
Cr$_{2}$O$_{3}$ & $R\overline{3}c$ (167)& 3.400 & 2.380 & 4.230 & 3.480 & N & \cite{tran2017}\\
Fe$_{2}$O$_{3}$ & $R\overline{3}c$ (167) & 2.200 & 1.400 & 3.320 & 2.490 & N & \cite{tran2017}\\
\bottomrule\end{longtable}

\bibliographystyle{apsrev4-2}\bibliography{references}